\documentclass[prb,twocolumn,preprintnumbers,amsmath,amssymb]{revtex4-1}

\usepackage{graphicx}
\usepackage{ulem}

\begin{document}
\title{ Pairing sizes in attractively interacting Fermi gases with Spin-orbit Couplings }
\author{ Yu Yi-Xiang,$^{1,2}$ Jinwu Ye,$^{2,3}$ and Wu-Ming Liu$^{1}$  }
\affiliation{
$^{1}$Beijing National Laboratory for Condensed Matter Physics,
Institute of Physics, Chinese Academy of Sciences,
Beijing 100190, China   \\
$^{2}$Department of Physics and Astronomy, Mississippi State University, MS, 39762, USA  \\
$^{3}$  Key Laboratory of Terahertz Optoelectronics, Ministry of Education,
       Department of Physics, Capital Normal University, Beijing, 100048, China  }

\date{\today }

\begin{abstract}
    Extensive research has been lavished on effects of spin-orbit couplings (SOC) in attractively interacting Fermi systems
    in both neutral cold atom systems and condensed matter systems.
    Recently, it was suggested that a SOC drives a new class of BCS to BEC crossover which is different than the conventional one without a SOC.
    Here, we explore what are the most relevant physical quantities to describe such a new BCS to BEC crossover and their experimental detections.
    We extend the concepts of the pairing length  and ``Cooper-pair size"  in the absence of SOC to Fermi systems with SOC.
    We investigate the dependence of chemical potential, pairing length, ``Cooper-pair size"
    on the SOC strength and the scattering length at  $ 3d $ (the bound state energy at $ 2d $) for three attractively interacting Fermi gases with
    3 dimensional (3d) Rashba, 3d Weyl and 2d Rashba SOC respectively.
    We show that only the pairing length can be used to characterize this new BCS to BEC crossover.
    Furthermore, it is the only length which can be directly measured by radio-frequency dissociation spectra type of experiments.
    We stress crucial differences among the pairing length, ``Cooper-pair size " and the two-body bound state size.
    Our results provide the fundamental and global picture of the new BCS to BEC crossover and  its experimental detections in various  cold atom and
    condensed matter systems.
\end{abstract}


\maketitle

\section{ Introduction}

 Spin-orbit coupling (SOC) has played important roles in various condensed matter systems such as anomalous Hall effects \cite{ahe},
 non-centrosymmetric superconductors with lifted spin degeneracy \cite{socsemi} and exciton superfluids in electron-hole
 semiconductor bilayers \cite{niu,wu}.
 Recently the investigation and control of spin-orbit coupling (SOC) have become subjects of
 intensive research after the discovery of the topological insulators \cite{Kane,Zhang}.
 For example, the SOC is a critical determining factor leading to a whole new class of electronic states  \cite{kitpconf} such as
 various spin-orbital ordered states, spin liquids, various topological phases, etc.
 The 1D SOC  which is a linear combination of Rashba and Dresselhauss SOC
 has been successfully generated in several experimental groups for neutral atoms
 both in Bose and Fermi gas \cite{SOCexp,SOCexp2,SOCexp3,gaugermp}.
 Possible experimental constructions of  2D  Rashba or Dresselhauss SOC and 3D Weyl SOC have also been proposed \cite{SOCtheo,ZFXu}.
 There are also extensive theoretical investigations on various important effects of SOC on
 attractively interacting \cite{twobody,crossover,zhai,hu} degenerate Fermi gases across BCS to BEC crossover.
 Collective excitations above the mean field states have also been calculated in \cite{collra2d,collra3d,colliso,colliso2}.
 The collective modes and magnetic transitions in repulsively interacting Fermi gas were investigated \cite{replusive1,replusive2}.
 Recently, both staggered \cite{stagg} and uniform \cite{uniform} artificial magnetic fields have been generated in optical lattices.
 Scaling functions for various gauge-invariant and non-gauge invariant quantities across topological transitions driven by the SOC on an optical lattice have been derived \cite{topo}. Especially, it was also stressed in \cite{topo} that in contrast to condensed matter experiments where only gauge invariant
 quantities can be measured, both gauge invariant and non-gauge invariant quantities can be measured by experimentally generating various non-Abelian gauges corresponding to the same set of Wilson loops.
 The interplay among the SOC, interactions and lattice geometries leading to new quantum phase, excitation spectrum and quantum phase transitions
 has also been explored \cite{rh}.

 The BCS to BEC crossover is a long outstanding problem in both condensed matter \cite{superbook,legg} and cold atoms \cite{crossrmp}.
 Conventional superconductors is well side the BCS limit, so mean field theory works well \cite{superbook,legg}.
 Due to its short coherence $ \xi $, high temperature superconductors are near the BCS to BEC crossover, but still in the BCS side
 with well defined Fermi surface \cite{legg,hightc}, so quantum fluctuation effects are large.
 The BCS to BEC crossover of exciton superfluids in electron hole semiconductor bilayer can be
 tuned by the exciton density \cite{power,squ,excitoncorr,exciton2}. The BCS to BEC crossovers of attractively
 interacting neutral fermions are tuned by sweeping across a Feshbach resonance \cite{crossrmp}.
 The effects of SOC on the BCS to BEC  crossover has been investigated by several groups \cite{twobody,crossover,zhai,hu}.
 Especially, the authors in \cite{crossover} found that when the SOC strength is well beyond the value of the topological Lifshitz transition
 of the non-interacting  Fermi surfaces \cite{replusive2,topo}, the overlap between the two-body wavefunction \cite{twobody,superbook} and the
``Cooper-pair wavefunction " \cite{explain}  (see Sec. II-B for its definition)  approaches to be 1.
 So they concluded that at a fixed scattering length at the BCS side (in the absence of SOC),
 the SOC drives a new BCS to BEC crossover  which is in a different class
 than the one without a SOC driven by the exciton density \cite{power,squ,excitoncorr,exciton2} or Feshbach resonance \cite{crossrmp}.
 However, as to be shown in this paper, the ``Cooper-pair wavefunction" is useful for illustration purposes only instead of being physical,
 so its overlap with the two-body wavefunction is not physical, can not be measured experimentally.
 In order to describe the new BCS to BEC crossover driven by the SOC strength, it is important
 to identify and compute the most relevant physical quantity to describe such a new BCS to BEC crossover and
 then address its experimental detections.


  In this paper, we address this outstanding problem by investigating 3 SOC coupled Fermi gases:
  (1) a 3D Fermi gas with a Rashba SOC (2) a 3D Fermi gas with a isotropic Weyl SOC
  (3) a 2D Fermi gas with a Rashba SOC.
  We first extend the concepts of the pairing length associated with a many-body wavefunction \cite{superbook,legg,melo}
  and ``Cooper-pair size" \cite{explain} associated with a ``Cooper-pair  wavefunction" in the absence of SOC to Fermi systems with a SOC.
  The three systems have different symmetries: the $ [ U(1)_{spin} \times U(1)_{orbit} ]_D $ symmetry at 3d where the $ D $ means the spontaneous rotation in spin and orbital space, the $ [ SU(2)_{spin} \times SO(3)_{orbit} ]_D $ symmetry at 3d
  and  the $ [ U(1)_{spin} \times U(1)_{orbit} ]_D $ symmetry at 2d respectively.
  These symmetries determine the number of independent pairing lengths and Cooper-pair sizes to be 2, 1 and 1 respectively.
  We then study the dependence of chemical potential, pairing length, ``Cooper-pair size"
 on the SOC strength $ \lambda $ for three attractively interacting Fermi gases.
  We show that from the BCS side at $ \lambda=0 $,  as the SOC strength increases, the chemical potential
  drops below the bottom of the single particle spectrum $ \mu_0= - \frac{ \lambda^{2}}{2m} $,
  so it can be used to characterize qualitatively the BCS to BEC crossovers driven by the SOC strength.
    The pairing length decreases monotonically and quickly below the inter-particle spacing, so
    can be used to characterize quantitatively the BCS to BEC crossovers. Furthermore, the pairing length can be directly measured by
    using radio-frequency dissociation spectra type of experiments \cite{mitdiss} as soon as the 2d and 3d SOC can be realized experimentally.
    In sharp contrasts, the ``Cooper-pair size" used in the previous work  \cite{twobody,crossover,zhai} shows non-monotonic behaviors, so it may not be used to
    characterize the BCS to BEC crossover even qualitatively.   Furthermore, it is not an experimentally measurable quantity.
   Starting from the BEC side at $ \lambda=0 $, the effects of SOC are small, both the pairing length and the ``Cooper-pair size" converge to
    the two-body bound state size \cite{twobody,superbook}.
  We conclude that the pairing length is a much more robust concept than the ``Cooper pair size ", it
  is also the only experimentally detectable physical quantity which can be used to describe the BCS to BEC crossover even quantitatively.
 We also discuss relations among the many-body BCS wavefunctions, the ``Cooper-pair wavefunctions" and the two-body wavefunctions,
  therefore stress crucial differences among the pairing length, Cooper-pair size and the two body bound state size.
  The results provide a solid foundation for the fundamental physics of the new BCS to BEC crossover and  its experimental detections.
  Our results should also shed considerable lights on condensed matter systems such as 2d exciton superfluids and 2d non-centrosymmetric superconductors.

   The rest of the paper is organized as follows: In Sect. II, we first review the different definitions and concepts of pairing length and
   ``Cooper pair size", then extend their concepts to SOC system where the spin is not a conserved quantity.
   In Sect.III, for a 3d Rashba systems, we study how the chemical potential, the pairing length and the Cooper pair size change as the SOC strength increases,
   especially focus on their  behaviors across the new BCS to BEC crossover driven by the SOC strength.
   To be complete, we also study how the the pairing length and the Cooper pair size change with the scattering lengths at fixed SOC strengths.
   We also discuss the crucial differences among the pairing length, Cooper-pair size and the two-body bound state size.
   In Sec.IV,  we compute the same quantities on 3d Fermi gas with an isotropic Weyl SOC.
   In Sec.V, we perform similar calculations on 2d Fermi gas with Rashba SOC
   which needs a different regularization than the two 3d systems discussed in the previous two sections. In Sec.VI, we discuss the implications of the results
   achieved in the previous sections, especially in the Sec. V, on several condensed matter systems.
   We summarize the main results and discuss several exciting perspectives in Sec. VII.

\section{ Extend paring length and Cooper pair size to SOC systems }

We consider a homogeneous Fermi gas with an attractive contact potential:
\begin{eqnarray}
H &=&\underset{\boldsymbol{p},\sigma =\uparrow ,\downarrow }{\sum }c_{%
\boldsymbol{p}\sigma }^{\dagger }\left(\frac{\boldsymbol{p}^{2}}{2m}-\mu
\right) c_{\boldsymbol{p}\sigma }+V_{soc}  \nonumber \\
&&+\frac{g}{V}\underset{\boldsymbol{p},\boldsymbol{q},\boldsymbol{s}}{\sum }%
c_{\frac{\boldsymbol{s}}{2}+\boldsymbol{p}\uparrow }^{\dagger }c_{\frac{%
\boldsymbol{s}}{2}-\boldsymbol{p}\downarrow }^{\dagger }c_{\frac{\boldsymbol{%
s}}{2}-\boldsymbol{q}\downarrow }c_{\frac{\boldsymbol{s}}{2}+\boldsymbol{q}%
\uparrow },
\label{hfermion}
\end{eqnarray}%
where $d=2,3$, and $ V_{soc}$ is the spin-orbit coupling (SOC) term which can
be Rashba or Dresselhauss type or 3d Weyl isotropic  SOC.

It was known that the interaction $ g $ need to be regularized differently
in 2d and 3d. In 3d, the $ g $ can be
regularized by the s-wave scattering length $a_{s}$: $\frac{1}{g}=\frac{m}{%
4\pi \hbar ^{2}a_{s}}-\frac{1}{V}\underset{\boldsymbol{p}}{\sum }\frac{1}{%
2\epsilon _{\boldsymbol{p}}}$ where $V$ is the volume of the system and $%
\epsilon _{\boldsymbol{p}}=\frac{\boldsymbol{p}^{2}}{2m}$ is the free particle dispersion.
In 2d, the $ g $ can be regularized by the two-body binding energy $%
\epsilon _{B}$: $\frac{1}{g}=-\frac{1}{V}\underset{\boldsymbol{p}}{\sum }%
\frac{1}{2\epsilon _{\boldsymbol{p}}+\epsilon _{B}}$.

By introducing the order parameter $\Delta =\frac{g}{V}\underset{\boldsymbol{p}}{\sum }%
\left\langle c_{-\boldsymbol{p}\downarrow }c_{\boldsymbol{p}\uparrow
}\right\rangle $, one can reduce the interaction term to the mean-field
form: $H_{int}^{MF}=\underset{\boldsymbol{p}}{\sum }\left(
\Delta^{*} c_{-\boldsymbol{p}\downarrow }c_{\boldsymbol{p}\uparrow }+ \Delta c_{%
\boldsymbol{p}\uparrow }^{\dagger }c_{-\boldsymbol{p}\downarrow }^{\dagger
} \right) -\frac{V\left\vert \Delta \right\vert ^{2}}{g}$. The
chemical potential $\mu $ and the order parameter $\Delta $ can be
determined by two self-consistent equations---the number equation: $n=\frac{1%
}{V}\underset{\boldsymbol{p,\sigma }}{\sum }\left\langle c_{\boldsymbol{p}%
\sigma }^{\dagger }c_{\boldsymbol{p}\sigma }\right\rangle _{MF} $ and the
gap equation: $\Delta =\frac{g}{V}\underset{\boldsymbol{p}}{\sum }%
\left\langle c_{-\boldsymbol{p}\downarrow }c_{\boldsymbol{p}\uparrow
}\right\rangle _{MF}$.

   Without SOC, the pairing length has been calculated in Fermi gas across the whole BCS to BEC crossover tuned by Feshbach resonance in \cite{melo}.
   Most importantly, it has been measured in MIT's group lead by Ketterle using radio-frequency dissociation spectra throughout the whole BCS to BEC crossover in \cite{mitdiss}. However, the effects of SOC on the pairing correlation lengths have never been studied so far.
   In this section, we first review the definition and concepts of the  pairing length  without SOC, then extend  to the SOC case.

\subsection{ Pairing length }

   For a spin singlet superfluid without SOC, the fermion pair correlation functions are defined as:
\begin{equation}
   \psi(\vec{r})= \frac{1}{n^2} \langle c^{\dagger}_{\uparrow}(\vec{r}) c^{\dagger}_{\downarrow}(0) c_{\downarrow}(0) c_{\uparrow}(\vec{r}) \rangle
   - \frac{1}{4},
\label{gg}
\end{equation}
   where $ n $ is the particle density and the average is taken with respect to
   the BCS ground state  $ |\Phi \rangle = | BCS \rangle $ (in second quantized form):
\begin{eqnarray}
     | \Phi \rangle & = & \Pi_{\vec{k}} (u_{\vec{k}} + v_{\vec{k}} c^{\dagger}_{\vec{k} \uparrow} c^{\dagger}_{-\vec{k} \downarrow})|0 \rangle
            \nonumber   \\
     & = &
   (\Pi_{\vec{k}}  u_{\vec{k}}) \Pi_{\vec{k}}(1 + \frac{ v_{\vec{k}} }{ u_{\vec{k}} } c^{\dagger}_{\vec{k} \uparrow} c^{\dagger}_{-\vec{k} \downarrow})|0 \rangle    \nonumber   \\
    &= &(\Pi_{\vec{k}}  u_{\vec{k}})
    exp [ \sum_{\vec{k}} \frac{ v_{\vec{k}} }{ u_{\vec{k}} } c^{\dagger}_{\vec{k} \uparrow} c^{\dagger}_{-\vec{k} \downarrow})] |0 \rangle,
\label{bcswave0}
\end{eqnarray}
     which obviously hosts in-definite number of electrons. Its first quantized form was discussed in \cite{superbook}.

     The pairing length is defined as  \cite{melo,mitdiss}
\begin{equation}
   \xi^{2}_{pair}=\frac{ \int d \vec{r} \psi(\vec{r}) r^2 }{  \int d \vec{r} \psi(\vec{r}) }.
\label{pairr}
\end{equation}

   At the mean field level, Eqn.\ref{gg} reduces to:
\begin{equation}
   \psi(\vec{r})= \frac{1}{n^2} |\langle \Phi |c^{\dagger}_{\uparrow}(\vec{r}) c^{\dagger}_{\downarrow}(0)| \Phi \rangle|^{2},
\label{ggmean}
\end{equation}
    where $ |\Phi \rangle = | BCS \rangle $.

   Under the mean field approximation, Eqn.\ref{pairr} can be rewritten as \cite{mitdiss}:
\begin{equation}
   \xi^{2}_{pair}=\frac{ \langle \psi_{\alpha \beta} |r^{2} | \psi_{\alpha \beta} \rangle  }
   { \langle \psi_{\alpha \beta} | \psi_{\alpha \beta} \rangle  },
\label{pairr2}
\end{equation}
    where  $ \psi_{\alpha \beta}(\vec{r})= \langle \Phi | c^{\dagger}_{\alpha}(\vec{r}) c^{\dagger}_{\beta}(0)| \Phi \rangle $ with $ \alpha= \uparrow, \beta=\downarrow $.

    The Fourier transform of Eqn.\ref{pairr2} to $ \vec{k} $ space leads to:
\begin{equation}
   \xi^{2}_{pair}= \frac{ \langle \psi_{\alpha \beta} | \nabla^{2}_{\vec{k}} | \psi_{\alpha \beta} \rangle  }
   { \langle \psi_{\alpha \beta} | \psi_{\alpha \beta} \rangle  },~~~~\alpha= \uparrow, \beta=\downarrow,
\label{pairk}
\end{equation}
    where $ \psi_{\alpha \beta}(\vec{k})= \langle \Phi |c^{\dagger}_{\alpha}(\vec{k}) c^{\dagger}_{\beta}(-\vec{k})| \Phi \rangle $
    is the Fourier transform of $ \psi_{\alpha \beta}(\vec{r}) $ with $ \alpha= \uparrow, \beta=\downarrow $.
    More straightforwardly, Eqn.\ref{pairr2} in real space and Eqn.\ref{pairk} in momentum space are Fourier transform to each other.

    For a BEC to BCS crossover without SOC, the only pairing is the singlet pairing so
    $ \psi_{\uparrow \downarrow}(\vec{k})= \langle \Phi |c^{\dagger}_{\uparrow}(\vec{k}) c^{\dagger}_{\downarrow}(-\vec{k})| \Phi \rangle
    = u_{\vec{k}}v_{\vec{k}} = \frac{ \Delta_{0} }{ 2 E_{\vec{k} } } $ which is given  and shown in Fig.3 in \cite{melo}.

    It is important to point out that Eqn.\ref{gg} and Eqn.\ref{pairr} hold in general, while Eqn.\ref{ggmean} and \ref{pairr2} hold only at the mean field
    level. Only at the mean field level, one can ``intuitively" interpret Eqn.\ref{ggmean} and \ref{pairr2} as the expectation value
    of $ r^2 $ over the ``pairing wavefunction" $ \psi_{\alpha \beta}(\vec{r})= \langle \Phi | c^{\dagger}_{\alpha}(\vec{r}) c^{\dagger}_{\beta}(0)| \Phi \rangle $ with $ \alpha= \uparrow, \beta=\downarrow $. Although the concept of pairing length Eqn.\ref{pairr} hold in general,
    such a wavefunction interpretation breaks down beyond the mean field.

   In the presence of SOC, due to the non-conservation of spins, one need to average over all the spin components to define
   the fermion pair correlation functions, so Eqn.\ref{gg} should be replaced by
\begin{equation}
   \psi(\vec{r})= \frac{1}{n^2} \sum_{\alpha, \beta} \langle c^{\dagger}_{\alpha}(\vec{r}) c^{\dagger}_{\beta}(0) c_{\beta}(0) c_{\alpha}(\vec{r}) \rangle-1,
\label{ggspin}
\end{equation}
   where $ n $ is the particle density. Eqn.\ref{pairr} remains.

   At mean field level, following the steps to derive Eqns.\ref{ggmean} and \ref{pairr2} leads to
   the pairing length in the presence of SOC:
\begin{equation}
   \xi^{2}_{i} =  \frac{ \sum_{\vec{k}, \alpha, \beta} \langle \psi_{\alpha \beta} | \partial^{2}_{k_i} | \psi_{\alpha \beta} \rangle  }
   { \sum_{\vec{k}, \alpha, \beta} \langle \psi_{\alpha \beta} | \psi_{\alpha \beta} \rangle  },~~~~\alpha, \beta=\uparrow,\downarrow,
\label{xii}
\end{equation}
    which could be measured by radio-frequency dissociation spectra used in the experiment \cite{mitdiss} in the presence of SOC.
    After the spin sum, the orbital symmetry of the $ U(1)_{orbit}, O(3)_{orbit},  U(1)_{orbit} $ of the three systems to be discussed in the following
    three sections will be recovered.
    However, one still need to distinguish the pairing
    correlation length within $ xy $ plane and along $ z $ direction $ \xi^{2}_{xy} \neq  \xi^{2}_{z} $ in the first and the third system.


\subsection{ ``Cooper pair size" }

    In fact, one can also define the ``Cooper-pair size" through the ``Cooper pair wavefunction" \cite{superbook,read}.
    Removing the exponential in the normalized BCS wave function without SOC in Eqn.\ref{bcswave0} leads to the singlet ``Cooper-pair wavefunction" in second quantized form:
\begin{equation}
  | g_{cp} \rangle= \sum_{\vec{k}} \frac{ v_{\vec{k}} }{ u_{\vec{k}} } c^{\dagger}_{\vec{k} \uparrow} c^{\dagger}_{-\vec{k} \downarrow} |0 \rangle,
\end{equation}
    which hosts only two paired electrons. It can be understood as the two electron components of the many-body wavefunction.

    One can extract the ``Cooper-pair wavefunction" in the real space in the first quantization:
\begin{equation}
    g_{cp}(\vec{r})  =  g_{\uparrow \downarrow}(\vec{r}) (|\uparrow \downarrow \rangle - |\downarrow \uparrow  \rangle),~~
    g_{\uparrow \downarrow}(\vec{r}) =  \sum_{\vec{k}} e^{ i \vec{k} \cdot \vec{r} } \frac{ v_{\vec{k}} }{ u_{\vec{k}} }.
\label{singlet0}
\end{equation}
    It is necessary to point out that this ``Cooper-pair wavefunction" is different than the original
    pairing problem of two electrons near a Fermi surface first achieved by Cooper by solving
    the Schrodinger equation \cite{superbook}.

    The ``Cooper pair size" \cite{explain} is defined by \cite{superbook,read}
\begin{equation}
   l^{2}_{pair}=\frac{ \int d \vec{r} | g_{\uparrow \downarrow}(\vec{r}) |^{2} r^2 }{  \int d \vec{r} | g_{\uparrow \downarrow}(\vec{r}) |^{2} }
   = \frac{\left\langle g_{cp} \right\vert r^{2}\left\vert g_{cp}
\right\rangle }{\left\langle g_{cp} | g_{cp} \right\rangle }.
\label{pairrcp}
\end{equation}

    The Fourier transform of Eqn.\ref{pairrcp} to $ \vec{k} $ space leads to:
\begin{equation}
   l^{2}_{pair}= \frac{ \langle g_{\alpha \beta} | \nabla^{2}_{\vec{k}} | g_{\alpha \beta} \rangle  }
   { \langle g_{\alpha \beta} | g_{\alpha \beta} \rangle  },~~~~\alpha= \uparrow, \beta=\downarrow,
\label{pairkcp}
\end{equation}
    where $ g_{\uparrow \downarrow}(\vec{k})= \frac{ v_{\vec{k}} }{ u_{\vec{k}} } $
    is the Fourier transform of $ g_{\uparrow \downarrow}(\vec{r}) $.
    This should be contrasted with $ \psi_{\uparrow \downarrow}(\vec{k})= \langle \Phi |c^{\dagger}_{\uparrow}(\vec{k}) c^{\dagger}_{\downarrow}(-\vec{k})| \Phi \rangle  = u_{\vec{k}}v_{\vec{k}} = \frac{ \Delta_{0} }{ 2 E_{\vec{k} } } $ used in Eqn.\ref{pairk}.

    It is important to point out that Eqn.\ref{bcswave0},\ref{singlet0},\ref{pairrcp},\ref{pairkcp} hold only in mean field
    level. Only at the mean field level, one can ``intuitively" interpret Eqn.\ref{pairrcp} and \ref{pairkcp} as the expectation value
    of $ r^2 $ over the ``Cooper-pair wavefunction" Eqn.\ref{singlet0}. However, the concept of Cooper-pair size Eqn.\ref{pairkcp}
    breaks down beyond the mean field theory.

    In the presence of SOC, after writing the mean field ground state in
    the form $ | BCS \rangle_{SOC} \propto  exp [ \sum_{\vec{k}} g_{\alpha \beta}(\vec{k}) c^{\dagger}_{\vec{k} \alpha } c^{\dagger}_{-\vec{k} \beta} ] |0 \rangle $, then one can extract the Cooper pair size  \cite{superbook,read} as
\begin{equation}
      l^{2}_{i} =  \frac{ \sum_{\vec{k}, \alpha, \beta} \langle g_{\alpha \beta} | \partial^{2}_{k_i} | g_{\alpha \beta} \rangle  }
   { \sum_{\vec{k}, \alpha, \beta} \langle g_{\alpha \beta} | g_{\alpha \beta} \rangle  },~~~~\alpha, \beta=\uparrow,\downarrow,
\label{li}
\end{equation}
     where the average over all the spin components is performed.
     Note that $ g_{\alpha \beta}(\vec{k}) $ is very much different than $ \phi_{\alpha \beta}(\vec{k}) $,
     so we may expect quite different behaviors from the two lengths. These will be explicitly demonstrated
     in the following sections. It is the pairing length which is measured in the MIT experiment \cite{mitdiss}.


In the following, we apply the formalism to the 3d Rashba  SOC, 3d Weyl SOC and
2d Rashba SOC systems.

\section{ 3D Fermi gas with a Rashba SOC}

\begin{figure}[tbp]
\includegraphics[width=8cm]{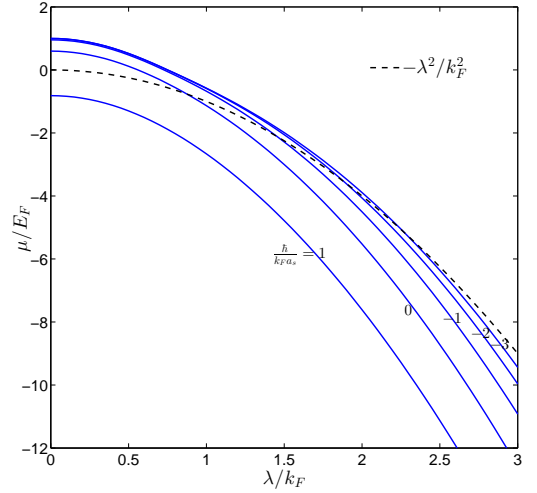}
\caption{ The chemical potential $\protect\mu $ versus $
\protect\lambda $ in a 3D Rashba SOC  for different scattering lengths.
The dashed line is the bottom of the single particle spectrum
$\protect\mu _{0}=-\frac{\protect\lambda ^{2}}{2m} $.
Starting from the BCS side at $ \lambda=0 $, as $\protect\lambda $ increases,
the $ \mu $ drops below  $ \protect\mu _{0}$. This fact indicates that the system evolves into
the BEC state. }
\label{fig1}
\end{figure}

 The 3d Rashba SOC can be written as%
\begin{equation}
V_{3d-ra}=\frac{\lambda }{m}\underset{\boldsymbol{p}}{\sum }p_{\perp }\left[
e^{-i\varphi _{\boldsymbol{p}}}c_{\boldsymbol{p}\uparrow }^{\dagger }c_{%
\boldsymbol{p}\downarrow }+e^{i\varphi _{\boldsymbol{p}}}c_{\boldsymbol{p}%
\downarrow }^{\dagger }c_{\boldsymbol{p}\uparrow }\right]  \label{3drashba},
\end{equation}%
where $\lambda $\ is the SOC strength, $p_{\perp }=\sqrt{%
p_{x}^{2}+p_{y}^{2}}$ and $\varphi _{\boldsymbol{p}}=Arg\left(
p_{x}+ip_{y}\right) $. This model has been studied by previous works \cite%
{zhai,hu} with different focuses. The single particle part $ H_0 $ in the Hamiltonian Eq. \ref{hfermion} can
be diagonalized in the helicity base as
\begin{equation}
H_{0}=\underset{\boldsymbol{p}}{\sum }%
\left(\xi _{\boldsymbol{p}+}h_{\boldsymbol{p}+}^{\dagger }h_{\boldsymbol{p}%
+}+\xi _{\boldsymbol{p}-}h_{\boldsymbol{p}-}^{\dagger }h_{\boldsymbol{p}
-}\right),
\label{h03dra}
\end{equation}
 where $\xi _{\boldsymbol{p}\pm }=\frac{p^{2}}{2m}\pm \frac{%
 \lambda p_{\perp }}{m}-\mu $ and the two helicity operators are:
\begin{eqnarray}
 h_{\boldsymbol{p}+} & = & \left[ c_{\boldsymbol{p}\uparrow
}+e^{-i\varphi _{\boldsymbol{p}}}c_{\boldsymbol{p}\downarrow }\right]/\sqrt{2},
     \nonumber   \\
h_{\boldsymbol{p}-} & = & \left[ e^{i\varphi _{\boldsymbol{p}%
}}c_{\boldsymbol{p}\uparrow }-c_{\boldsymbol{p}\downarrow }\right]/\sqrt{2}.
\label{helicitybase}
\end{eqnarray}.

In the helicity base, the mean-field interaction can be rewritten as $H_{int}^{MF}=-%
\frac{1}{2}\underset{\boldsymbol{p}}{\sum }\left(\Delta_0 e^{-i\varphi _{%
\boldsymbol{p}}}h_{\boldsymbol{p}+}^{\dagger }h_{\boldsymbol{-p}+}^{\dagger
}+\Delta_0 e^{i\varphi _{\boldsymbol{p}}}h_{\boldsymbol{p}-}^{\dagger }h_{%
\boldsymbol{-p}-}^{\dagger }+h.c.\right) -\frac{V\left\vert \Delta
\right\vert ^{2}}{g}$. The total Hamiltonian  $ H= H_0 + H_{int}^{MF} $ can be diagonalized by a
Bogoliubov transformation:%
\begin{eqnarray}
H &=&\underset{\boldsymbol{p}}{\sum }\left(E_{\boldsymbol{p}+}\alpha _{%
\boldsymbol{p}+}^{\dagger }\alpha _{\boldsymbol{p}+}+E_{\boldsymbol{p}%
-}\alpha _{\boldsymbol{p}-}^{\dagger }\alpha _{\boldsymbol{p}-}\right) \nonumber   \\
& - & %
\underset{\boldsymbol{p}}{\sum }\frac{E_{\boldsymbol{p}+}+E_{\boldsymbol{p}-}%
}{2}-\frac{V\left\vert \Delta_0 \right\vert ^{2}}{g},  \label{3drashbah}
\end{eqnarray}%
where the quasiparticle excitation energy $ E_{\boldsymbol{p}\pm }=\sqrt{\xi
_{\boldsymbol{p}\pm }^{2}+\left\vert \Delta_0 \right\vert ^{2}}$, and the quasi-particle
operators:%
\begin{eqnarray}
\alpha _{\boldsymbol{p}+} &=&\sqrt{\frac{E_{\boldsymbol{p}+}+\xi _{%
\boldsymbol{p}+}}{2E_{\boldsymbol{p}+}}}e^{i\varphi _{\boldsymbol{p}}}h_{%
\boldsymbol{p}+}-\sqrt{\frac{E_{\boldsymbol{p}+}-\xi _{\boldsymbol{p}+}}{2E_{%
\boldsymbol{p}+}}}h_{-\boldsymbol{p}+}^{\dagger },  \nonumber \\
\alpha _{\boldsymbol{p}-} &=&\sqrt{\frac{E_{\boldsymbol{p}-}+\xi _{%
\boldsymbol{p}-}}{2E_{\boldsymbol{p}-}}}e^{-i\varphi _{\boldsymbol{p}}}h_{%
\boldsymbol{p}-}-\sqrt{\frac{E_{\boldsymbol{p}-}-\xi _{\boldsymbol{p}-}}{2E_{%
\boldsymbol{p}-}}}h_{-\boldsymbol{p}-}^{\dagger },  \label{3drashbab}
\end{eqnarray}%
where all anticommutation relations hold ($\left\{ \alpha _{\boldsymbol{p}%
+},\alpha _{\boldsymbol{p}+}^{\dagger }\right\} =1$, $\left\{ \alpha _{%
\boldsymbol{p}+},\alpha _{\boldsymbol{p}-}^{\dagger }\right\} =0$, and so
on).

At zero temperature, the two self-consistent equations become
\begin{eqnarray}
n &=&\frac{1}{V}\underset{\boldsymbol{p}}{\sum }\left[ 1-\frac{\xi _{%
\boldsymbol{p}+}}{2E_{p+}}-\frac{\xi _{\boldsymbol{p}-}}{2E_{p-}}\right] ,
\nonumber \\
\frac{1}{g} &=&-\frac{1}{4V}\underset{\boldsymbol{p}}{\sum }\left[ \frac{1}{%
E_{p+}}+\frac{1}{E_{p-}}\right],
\label{3dconsistent}
\end{eqnarray}%
where, as said in the Sec.II, the interaction strength $g$ can be regularized by the s-wave
scattering length $a_{s}$: $\frac{1}{g}=\frac{m}{4\pi \hbar ^{2}a_{s}}-\frac{%
1}{V}\underset{\boldsymbol{p}}{\sum }\frac{1}{2\epsilon _{\boldsymbol{p}}}$.
In the rest of the section, we will determine the chemical potential $ \mu $, the paring length $ \xi_i $ in Eqn.\ref{xii},
the Cooper pair size $ l_i $ in Eqn.\ref{li}. Finally we will compare our many body results with the corresponding
two-body results \cite{twobody}.

\subsection{ Chemical potential across BCS to BEC crossover }

 One can find the chemical potential $\mu $ by
solving Eqn.\ref{3dconsistent}. It is shown in Fig.1.
As a contrast, the minimum of the single-particle excitation energy $\mu _{0}=%
\underset{\boldsymbol{p}}{\text{Min}}\{\frac{p^{2}}{2m}-\frac{\lambda
p_{\perp }}{m}\}=-\frac{\lambda ^{2}}{2m}$  is also plotted in Fig.1.
We can qualitatively assign the region with $\mu >\mu _{0}$ as the BCS region and  $\mu <\mu
_{0}$ as the BEC region. As shown in Fig.1, starting from the BCS side at $ \lambda=0 $, as $ \lambda $ increases,
the $ \mu $ drops below $ \mu_0 $. Therefore, we conclude that Rashba SOC can drive a
crossover from BCS to BEC as first pointed out in \cite{crossover}.

\subsection{ Pairing length across BCS to BEC crossover }

\begin{figure}[tbp]
\includegraphics[width=8cm]{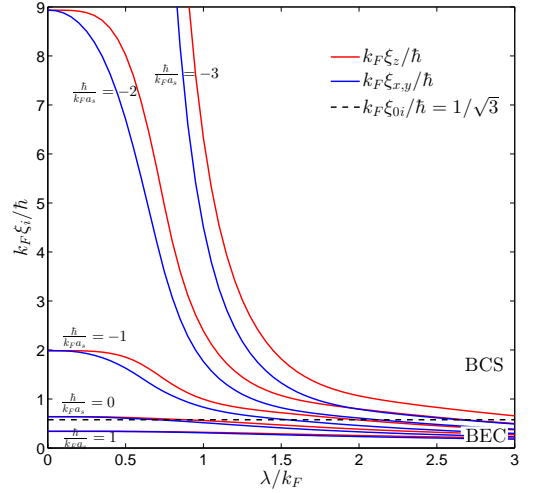} %
\caption{  The pairing length defined in Eqn.\ref{xii} along $ x $ direction (red lines) (
$\protect\xi _{x}=\protect\xi _{y}$) and  along the $z$ (blue lines) direction as a
functions of 3d Rashba SOC strength $\protect\lambda $. The dashed line is a guidance line
where $k_{F}\protect\xi_0 =\hbar $ ($k_F\protect\xi_{0i}=\hbar/%
\protect\sqrt{3}$ for each component). Starting from the BCS side at $ \lambda=0 $,
it decreases monotonically and quickly below the reference line, so describe precisely
the new BCS to BEC crossover driven by the SOC strength. Starting from the BEC side at $ \lambda=0 $, the effects of SOC are small. }
\label{fig2}
\end{figure}

\begin{figure}[tbp]
\includegraphics[width=8cm]{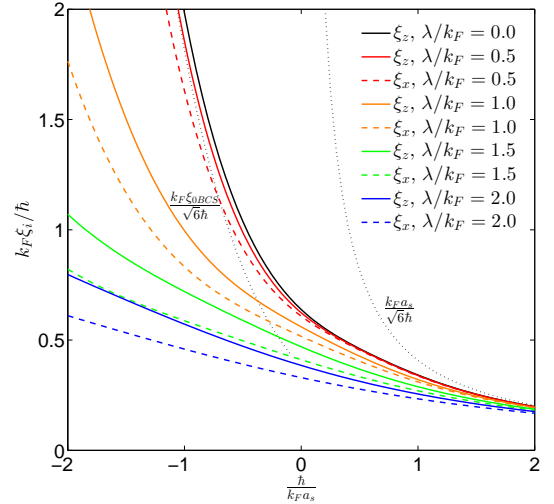}
\caption{ The pairing lengths $ \xi_z > \xi_x=\xi_y $ at a fixed 3d Rashba SOC strength versus the scattering length. Different colors stand for different SOC strengths.
Solid (dashed) lines stand for  $ \xi_z $ ($\xi_x=\xi_y $).
The dark dotted line on the left (right) is its BCS (BEC) limit $  \frac{1}{\sqrt{6}} k_F \xi_{0BCS}=\frac{1}{ 8 \sqrt{6} e^{2} } e^{-\frac{\pi \hbar }{2k_{F}a_{s}}} $ ($ \frac{ k_F a_s }{ \sqrt{6} } $) at $ \lambda=0 $. On the BCS side, the SOC effects are dramatic, but on the BEC side,
the SOC effects are small, all curves converge to the right dotted line  $ \frac{ k_F a_s }{ \sqrt{6} } $ from below. }
\label{fig3}
\end{figure}

The BCS ground state can be written as:%
\begin{eqnarray}
\left\vert BCS\right\rangle =\underset{\boldsymbol{p}}{\prod }^{\prime
}\alpha _{\boldsymbol{p}+}\alpha _{-\boldsymbol{p}+}\alpha _{\boldsymbol{p}%
-}\alpha _{-\boldsymbol{p}-}\left\vert 0\right\rangle
~~~~~~~~~~~~~~~~ \nonumber \\
\propto \exp \underset{\boldsymbol{p}}{\sum }^{\prime }\left[ w_{\boldsymbol{p}%
+}e^{-i\varphi _{\boldsymbol{p}}}h_{\boldsymbol{p}+}^{\dagger }h_{%
\boldsymbol{-p}+}^{\dagger }+w_{\boldsymbol{p}-}e^{i\varphi _{\boldsymbol{p}%
}}h_{\boldsymbol{p}-}^{\dagger }h_{\boldsymbol{-p}-}^{\dagger }\right]
\left\vert 0\right\rangle ,
\label{3drashbabcs}
\end{eqnarray}%
where the $ \prime $ means half of the momentum space and  $\left\vert 0\right\rangle $ is the electron vacuum state and $w_{%
\boldsymbol{p}\pm }=\sqrt{\frac{E_{\boldsymbol{p}\pm }-\xi _{\boldsymbol{p}%
\pm }}{E_{\boldsymbol{p}\pm }+\xi _{\boldsymbol{p}\pm }}}$.


   From Eqn.\ref{3drashbabcs}, one can find the singlet pairing amplitude:
\begin{eqnarray}
\psi _{\uparrow \downarrow }\left(\boldsymbol{p}\right) &=&\frac{%
\left\langle BCS\right\vert c_{\boldsymbol{p}\uparrow }^{\dagger }c_{-%
\boldsymbol{p}\downarrow }^{\dagger }\left\vert BCS\right\rangle }{%
\left\langle BCS|BCS\right\rangle } \\
&=&-\frac{1}{2}\left(\frac{w_{\boldsymbol{p}+}}{1+w_{\boldsymbol{p}+}^{2}}+%
\frac{w_{\boldsymbol{p}-}}{1+w_{\boldsymbol{p}-}^{2}}\right)   \nonumber   \\
&= & - \frac{ \Delta_{0} }{4} \sum_{\alpha=\pm} 1/E_{\vec{p},\alpha }   \nonumber   \\
&=&-\psi _{\downarrow \uparrow }\left(-\boldsymbol{p}\right) =-\psi
_{\downarrow \uparrow }\left(\boldsymbol{p}\right),
\label{3dsinglet}
\end{eqnarray}
      and the triplet pairing amplitude:
\begin{eqnarray}
\psi _{\uparrow \uparrow }\left(\boldsymbol{p}\right) &=&\frac{\left\langle
BCS\right\vert c_{\boldsymbol{p}\uparrow }^{\dagger }c_{-\boldsymbol{p}%
\uparrow }^{\dagger }\left\vert BCS\right\rangle }{\left\langle
BCS|BCS\right\rangle }  \nonumber \\
&=&\frac{1}{2}\left(\frac{w_{\boldsymbol{p}+}}{1+w_{\boldsymbol{p}+}^{2}}-%
\frac{w_{\boldsymbol{p}-}}{1+w_{\boldsymbol{p}-}^{2}}\right) e^{i\varphi _{%
\boldsymbol{p}}}   \nonumber  \\
& = & -  \frac{ \Delta_{0} }{ 4} e^{i \varphi_{\vec{p}} } \sum_{\alpha=\pm} \alpha/E_{\vec{p},\alpha }   \nonumber  \\
&=&-\psi _{\downarrow \downarrow }^{\ast }\left(\boldsymbol{p}\right).
\label{3dtriplet}
\end{eqnarray}

  Plugging into Eqn.\ref{xii} leads to the many body pairing length $ \xi_i $ along different directions versus the SOC strength shown in Fig.2.
We also plot a reference line  $k_{F}\xi _{0}=\hbar $
($k_{F}\xi _{0i}=\frac{\hbar }{\sqrt{3}}$ for each component) to qualitatively signal the
BCS to BEC crossover. As shown in the Fig.2, the pairing length in
both  $x$ (or $y$) and $ z $ direction decrease monotonically and sharply as the
SOC strength increases for a fixed interaction strength $\frac{\hbar }{%
k_{F}a_{s}}$, and finally drop below the reference line. This is the most direct
evidence that the Rashba SOC drives a crossover from BCS to BEC.
The monotonic decreasing shape of the pairing length in Fig.2 can be directly detected by radio-frequency dissociation spectra experiment \cite{mitdiss}.

  In the absence of the SOC when $\lambda =0$, there is only a singlet pairing,
  one can get an analytical result:
\begin{eqnarray}
\xi _{x,y,z}^{2}(\lambda=0)  &=& \frac{\hbar ^{2}4\int dpp^{d+1}\frac{\xi _{\boldsymbol{p}%
}^{2}}{E_{\boldsymbol{p}}^{6}}}{d\left(2m\right) ^{2}\int dp\frac{p^{d-1}}{%
E_{\boldsymbol{p}}^{2}}}  \nonumber \\
&=&\frac{\hbar ^{2}\sqrt{\frac{\eta \left(16\eta ^{4}+52\eta ^{2}+45\right)
}{\eta ^{2}+1}+\frac{16\eta ^{4}+44\eta ^{2}+25}{\sqrt{\eta ^{2}+1}}}}{%
\left(2m\Delta \right) 24\sqrt{\eta +\sqrt{\eta ^{2}+1}}},
\label{3drashbasize0}
\end{eqnarray}%
where $\eta =\frac{\mu }{\Delta }$.
In the weak coupling (BCS) limit\cite{melo} $
\frac{\hbar }{k_{F}a_{s}}\longrightarrow -\infty $, $\mu =E_{F}$ and $
\Delta_0 =\frac{8E_{F}}{e^{2}}e^{\frac{\pi \hbar }{2k_{F}a_{s}}
}\longrightarrow 0$, $\xi _{x,y,z}=\frac{1}{\sqrt{6}}\frac{
\hbar k_{F} }{2m\Delta_0} = \frac{1}{\sqrt{6}} \xi_{0BCS} $ where
$ \xi_{0BCS} = \frac{ \hbar v_F }{ 2 \Delta_0 } $ is nothing but the coherence length \cite{superbook} which
goes to $\infty $ as $\frac{\hbar
}{k_{F}a_{s}}\longrightarrow -\infty $.  In the strong coupling (BEC)
limit\cite{melo} $\frac{\hbar }{k_{F}a_{s}}\longrightarrow \infty $, $\mu =-\frac{E_b}
{ 2}+\frac{2k_{F}a_{s}}{3\pi \hbar }E_{F}$ where $ E_b= \frac{1}{ m a^{2}_{s} } $ is the binding energy, the
$\Delta_0 =\sqrt{\frac{16\hbar }{\pi k_{F}a_{s}}} E_{F}$,
so $\xi_{x,y,z} \rightarrow \frac{ a_s }{ \sqrt{6} } $ which recovers the two-body scattering length
Eqn.\ref{bound} \cite{factor}.

 The pairing lengths along different directions versus the scattering length is shown in Fig.3
 which is complementary to Fig.2.

\subsection{ Cooper pair size across BCS to BEC crossover }

\begin{figure}[tbp]
\includegraphics[width=8cm]{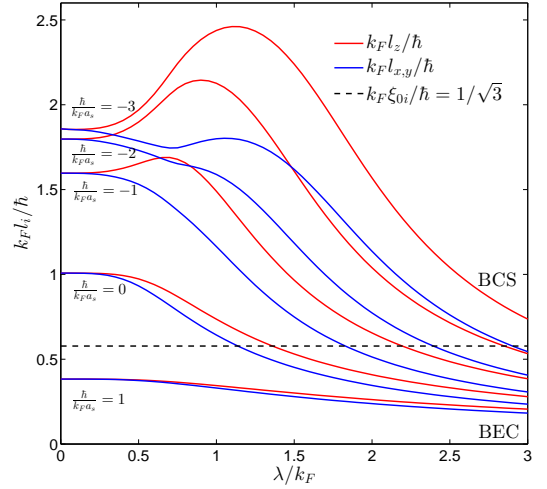}
\caption{  The Cooper pair size defined in Eqn.\ref{li}
\protect\ref{3drashbal} along $ x $ direction (red lines) ($l _{x}=l _{y}$) and $ z $ direction (blue
lines) as a function of 3d Rashba SOC strength $\protect\lambda $. Note its non-monotonic behaviors in the BCS side.
The effects of SOC are small starting from the BEC side at $ \lambda=0 $. }
\label{fig4}
\end{figure}

\begin{figure}[tbp]
\includegraphics[width=8cm]{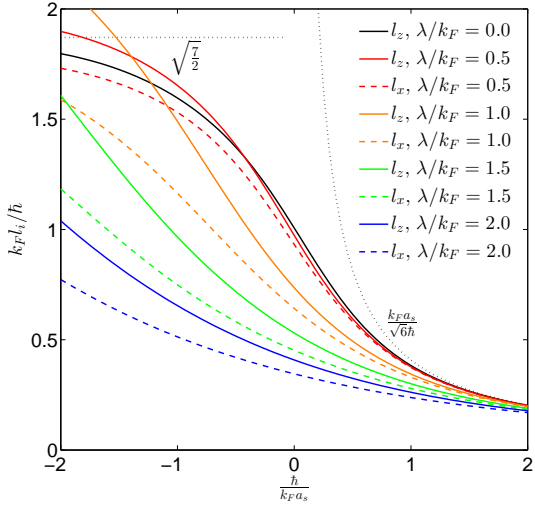}
\caption{ The Cooper pair size $ l_z > l_x=l_y $ at a fixed 3d Rashba SOC versus the scattering length. Different colors stand for different SOC strengths.
Solid (dashed) lines stand for  $ l_z $ ($ l_x=l_y $).
The dark dotted line on the left (right) is its BCS (BEC) limit $ \sqrt{7/2} $ ($ \frac{ k_F a_s }{ \sqrt{6} } $) at $ \lambda=0 $.
On the BCS side, the SOC effects are dramatic, but on the BEC side,
the SOC effects are small, all curves converge to the right dotted line  $ \frac{ k_F a_s }{ \sqrt{6} } $ from below.  }
\label{fig5}
\end{figure}

As shown in Sec.II, the Cooper pair size in Eqn.\ref{li} is another characteristic length in Fermi gas system.
Formally, one can define the ``Cooper pair wavefunction" \cite{crossover} in the second quantized form by removing
the exponential  in Eqn.\ref{3drashbabcs}:
\begin{eqnarray}
\left\vert g_{cp} \right\rangle &=&\underset{\boldsymbol{p}}{\sum }^{\prime }%
\left[ w_{\boldsymbol{p}+}e^{-i\varphi _{\boldsymbol{p}}}h_{\boldsymbol{p}%
+}^{\dagger }h_{\boldsymbol{-p}+}^{\dagger }+w_{\boldsymbol{p}-}e^{i\varphi
_{\boldsymbol{p}}}h_{\boldsymbol{p}-}^{\dagger }h_{\boldsymbol{-p}%
-}^{\dagger }\right] \left\vert 0\right\rangle  \nonumber \\
&=&\underset{\boldsymbol{p}}{\sum }^{\prime } [ g_{\uparrow \downarrow
}\left(\boldsymbol{p}\right) c_{\boldsymbol{p}\uparrow }^{\dagger }c_{%
\boldsymbol{-p}\downarrow }^{\dagger }+g_{\downarrow \uparrow }\left(
\boldsymbol{p}\right) c_{\boldsymbol{p}\downarrow }^{\dagger }c_{\boldsymbol{%
-p}\uparrow }^{\dagger }      \nonumber   \\
& & +  g_{\uparrow \uparrow }\left(\boldsymbol{p}\right)
c_{\boldsymbol{p}\uparrow }^{\dagger }c_{\boldsymbol{-p}\uparrow }^{\dagger
}+g_{\downarrow \downarrow }\left(\boldsymbol{p}\right) c_{\boldsymbol{p}%
\downarrow }^{\dagger }c_{\boldsymbol{-p}\downarrow }^{\dagger } ]
\left\vert 0\right\rangle,
\label{3drashbacp}
\end{eqnarray}%
  which only has two paired electrons with both singlet and triplet pairing.   In Eqn.\ref{3drashbacp}, we have used Eqn.\ref{helicitybase} and found:
\begin{eqnarray}
g_{\uparrow \downarrow }\left(\boldsymbol{p}\right) &=&-\frac{1}{2}\left(
w_{\boldsymbol{p}+}+w_{\boldsymbol{p}-}\right) =-g_{\downarrow \uparrow
}\left(-\boldsymbol{p}\right) =-g_{\downarrow \uparrow }\left(\boldsymbol{p%
}\right),  \nonumber \\
g_{\uparrow \uparrow }\left(\boldsymbol{p}\right) &=&\frac{1}{2}\left(w_{%
\boldsymbol{p}+}-w_{\boldsymbol{p}-}\right) e^{-i\varphi _{\boldsymbol{p}%
}}=-g_{\downarrow \downarrow }^{\ast }\left(\boldsymbol{p}\right).
\label{3drashbawf2}
\end{eqnarray}%

  The corresponding first quantized form of Eqn.\ref{3drashbacp} in real space is:
\begin{eqnarray}
 g_{cp}(\vec{r}) &= &  g_{\uparrow \downarrow }(\vec{r}) (| \uparrow \downarrow \rangle-  |  \downarrow \uparrow \rangle)
                               \nonumber   \\
 & + & g_{\uparrow \uparrow }(\vec{r})  | \uparrow \uparrow \rangle
 - g^{*}_{\uparrow \uparrow }(\vec{r}) | \downarrow \downarrow \rangle,
\label{3drashbacpfirst}
\end{eqnarray}
   where $ g_{\alpha \beta }(\vec{r})=\sum_{\vec{p}} e^{ i \vec{p} \cdot \vec{r} }
   g_{\alpha \beta}(\vec{p}), \alpha,\beta= \uparrow, \downarrow $.
   Compared to Eqn.\ref{singlet0}, one can see that there are two extra equal-spin $ p_x \pm i p_y $  pairing components\cite{legg} similar to
   the $ A $-phase of $^{3} He $.

   It is easy to see that the Cooper pair size along the $i$ direction in Eqn.\ref{li} can be expressed as:
\begin{equation}
l_{i}^{2}=\frac{\left\langle g_{cp} \right\vert r_{i}^{2}\left\vert g_{cp}
\right\rangle }{\left\langle g_{cp} |g_{cp} \right\rangle },~~~~i=x,y,z,
\label{3drashbal}
\end{equation}
  which has a clear physical meaning: the Cooper pair size is the ``average size" of the Cooper pair
  wavefunction Eqn.\ref{3drashbacpfirst}. Shown in Fig.4 is the
the Cooper pair size along different directions versus the SOC strength.
In sharp contrast to the pairing length, it is non-monotonic \cite{uncover} in the BCS side $ a_s < 0 $, so may not be a good quantity to characterize
the BCS to BEC crossover. Furthermore, it may not be an experimentally detectable quantity anyway.

In the absence of the SOC when $\lambda =0$, there is only a singlet pairing, Eqn.\ref{3drashbal} is simplified to:
\begin{equation}
l_{i}^{2}(\lambda=0) = \frac{\hbar ^{2}\underset{\boldsymbol{k}}{\sum }\left(1-\frac{\xi
_{\boldsymbol{k}}}{E_{\boldsymbol{k}}}\right) ^{2}\frac{k^{2}}{m^{2}}}{3%
\underset{\boldsymbol{k}}{\sum }\left(E_{\boldsymbol{k}}-\xi _{\boldsymbol{k%
}}\right) ^{2}}.  \label{3dnosocl}
\end{equation}%
In the weak coupling (BCS) limit (i.e.$\frac{\hbar }{k_{F}a_{s}}%
\longrightarrow -\infty $), $\mu =E_{F}$ and $\Delta \rightarrow 0$, one finds $
l_{x,y,z}\longrightarrow \sqrt{\frac{7}{2}}\frac{\hbar }{k_{F}}$ which is nothing but the inter-particle distance,
in sharp contrast to the pairing length which is nothing but the coherence length.
Using $ l_{0BCS} \sim \frac{ \hbar^2 k_F }{ 2 m \epsilon_F }, \xi_{0BCS} \sim  \frac{ \hbar^2 k_F }{ 2 m \Delta_0 } $,
one can see their ratio $ l_{0BCS}/\xi_{0BCS} \sim \Delta_0/\epsilon_F =\frac{8}{e^{2}}e^{\frac{\pi \hbar }{2k_{F}a_{s}}
}\longrightarrow 0 $. For conventional superconductors $  l_{0BCS}/\xi_{0BCS} \sim 10^{-4} $ which indicates that there are about
$ 10^4 $ other Cooper pairs inside a given Cooper pair. However, for high $ T_c $ superconductors \cite{legg,hightc},
$  l_{0BCS}/\xi_{0BCS} \sim 10^{-1} $ which indicates that they are quite close to the BCS to BEC crossover, but still in the BCS side.

In the BEC limit, we find $ l_i \rightarrow \frac{ a_s }{ \sqrt{6} } $ which also recovers the two-body scattering length
Eqn.\ref{bound}\cite{factor}. So $ l_{0BCS}/\xi_{0BCS}=1 $ in the strong  BEC limit. This should be expected because
the Cooper pair wavefunction is nothing but the two electrons component of the many-body wavefunction, so both lengths have to be the same in the
strong BEC limit.

The Cooper pair sizes along different directions versus the scattering length is shown in Fig.5 which is complementary to Fig.4

\subsection{ Contrast with two-body wavefunctions }

The 2-body wavefunction with a 3d Rashba SOC was worked out in \cite{twobody} by solving a 2-body Schrodinger equation \cite{superbook}.
It is instructive to compare the many-body wave functions Eqn.\ref{3drashbabcs} and the Cooper pair wavefunction Eqn.\ref{3drashbawf2}
with the corresponding two-body wave functions (see the extreme oblate case in \cite{twobody}).
They all have the same symmetries, namely:
\begin{eqnarray}
\psi_{\uparrow \downarrow }\left(\boldsymbol{p}\right) & = & -\psi _{\downarrow
\uparrow }\left(\boldsymbol{p}\right),~~~~ g_{\uparrow \downarrow }\left(
\boldsymbol{p}\right) =-g_{\downarrow \uparrow }\left(\boldsymbol{p}\right),    \nonumber   \\
\psi _{\uparrow \uparrow }\left(\boldsymbol{p}\right) & = & -\psi
_{\downarrow \downarrow }^{\ast }\left(\boldsymbol{p}\right),~~~~
g_{\uparrow \uparrow }\left(\boldsymbol{p}\right) =-g_{\downarrow
\downarrow }^{\ast }\left(\boldsymbol{p}\right).
\end{eqnarray}

However, they have quite different behaviors. It was shown that in the absence of the SOC when $\lambda =0$, there exist a bound state in the BEC side only with $ a_s > 0 $, the bound state has only a singlet component $ \psi_0(\vec{r})= \frac{1}{r} e^{-r/a_s} $ with a binding energy $ E_b=\frac{ \hbar^2 }{ m a_s } $.
It is easy to see the size of the bound state:
\begin{equation}
 b(\lambda=0) =\frac{\left\langle \psi_{0} \right\vert r ^{2}\left\vert \psi_{0}
\right\rangle }{\left\langle \psi_{0} |\psi_{0} \right\rangle }= \frac{ a_s }{\sqrt{2}},
\label{bound}
\end{equation}
   which is identical to both the pairing size Eqn.\ref{xii} and the Cooper-pair size Eqn.\ref{li} in the BEC limit \cite{factor}.
   As shown in Fig.1, in the absence of the SOC when $\lambda =0$,
   the pairing length $ \xi_i $ and Cooper pair size $ l_i $  are well defined in both the BCS and BEC limit.

However, as shown in \cite{twobody}, a non-zero SOC strength $ \lambda \neq 0 $ will always lead to a two-body bound state at any $ a_s $,
extending the $ b(\lambda=0) $ in Eqn.\ref{bound} to a non-zero $ \lambda $ can be easily calculated using
the two body wavefunctions in \cite{twobody,colliso}.
Any non-zero SOC strength, as shown in Fig.1, leads to $ \xi_z > \xi_x=\xi_y $ and  $ l_z > l_x=l_y $.
In the BCS side, $ \xi_i $ and $ l_i $ display dramatically different behaviors. While in the BEC limit
both $ \xi_i $ and $ l_i $ converge to the size of the two-body bound state.
This is expected, in the strong BEC limit, the overlap between the two-body wavefunction and the many-body wavefunction must be the same
as that between the two-body wavefunction and the Cooper-pair wavefunction.

\section{ 3D Fermi gas with an isotropic Weyl SOC}

\begin{figure}[tbp]
\includegraphics[width=8cm]{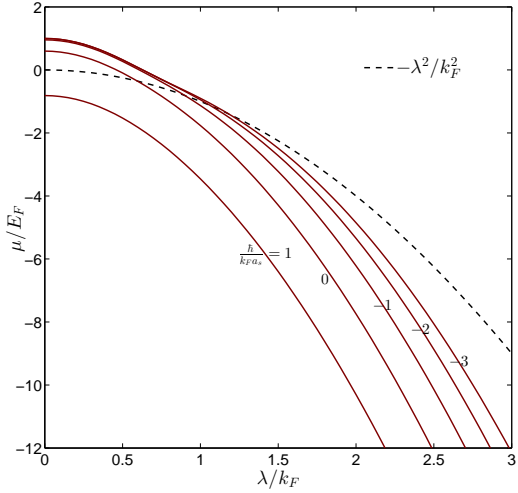} %
\caption{  The chemical potential $\protect\mu $ versus the  3D isotropic Weyl SOC strength $%
\protect\lambda $ for different scattering lengths. The black dashed line
$\protect\mu_{0}=-\frac{\protect\lambda ^{2}}{2m}$is the
chemical potential at the bottom of single particle spectrum.
Starting from the BCS side at $ \lambda=0 $, as $\protect\lambda $ increases,
the $ \mu $ drops below  $ \protect\mu _{0}$  indicating a crossover from BCS to BEC. }
\label{fig6}
\end{figure}

  The 3d Weyl SOC term can be written as%
\begin{eqnarray}
V_{Weyl} & = & \frac{\lambda }{m}\underset{\boldsymbol{p}}{\sum } [ p_{\perp
}\left(e^{-i\varphi _{\boldsymbol{p}}}c_{\boldsymbol{p}\uparrow }^{\dagger
}c_{\boldsymbol{p}\downarrow }+e^{i\varphi _{\boldsymbol{p}}}c_{\boldsymbol{p%
}\downarrow }^{\dagger }c_{\boldsymbol{p}\uparrow }\right)    \nonumber   \\
& + & p_{z}\left(c_{%
\boldsymbol{p}\uparrow }^{\dagger }c_{\boldsymbol{p}\uparrow }-c_{%
\boldsymbol{p}\downarrow }^{\dagger }c_{\boldsymbol{p}\downarrow }\right) ].
\label{3disotropic}
\end{eqnarray}%
The single particle part in Hamiltonian Eqn.\ref{hfermion} can be
diagonalized in the helicity bases as $H_{0}=\underset{\boldsymbol{p}}{\sum }%
\left(\xi _{\boldsymbol{p}+}h_{\boldsymbol{p}+}^{\dagger }h_{\boldsymbol{p}%
+}+\xi _{\boldsymbol{p}-}h_{\boldsymbol{p}-}^{\dagger }h_{\boldsymbol{p}%
-}\right) $ where $\xi _{\boldsymbol{p}\pm }=\frac{p^{2}}{2m}\pm \frac{%
\lambda p}{m}-\mu $ and the helicity operators:
\begin{eqnarray}
h_{\boldsymbol{p}+} & = & \sqrt{\frac{1}{2}}\left[ \sqrt{\frac{p+p_{z}}{p}}c_{%
\boldsymbol{p}\uparrow }+\sqrt{\frac{p-p_{z}}{p}}e^{-i\varphi _{\boldsymbol{p%
}}}c_{\boldsymbol{p}\downarrow }\right],  \nonumber  \\
h_{\boldsymbol{p}-} & = & \sqrt{%
\frac{1}{2}}\left[ \sqrt{\frac{p-p_{z}}{p}}e^{i\varphi _{\boldsymbol{p}}}c_{%
\boldsymbol{p}\uparrow }-\sqrt{\frac{p+p_{z}}{p}}c_{\boldsymbol{p}\downarrow
}\right].
\end{eqnarray}

 In the mean-field theory, the total Hamiltonian can also be
diagonalized as Eqn.\ref{3drashbah}, and the quasiparticle excitation
energy $E_{\boldsymbol{p}\pm }=\sqrt{\xi _{\boldsymbol{p}\pm
}^{2}+\left\vert \Delta_0 \right\vert ^{2}}$, and the Bogoliubov quasi-particle operators take
the same form as Eqn.\ref{3drashbab}. At zero temperature, the two
self-consistent equations also take the same form as Eqn.\ref%
{3dconsistent} with the corresponding  $\xi _{\boldsymbol{p}\pm }$ and $E_{\boldsymbol{p}\pm }$
defined above. Solving them leads to the chemical potential shown in Fig.6.
Similar to Fig.1, the Weyl SOC also drives a new crossover from BCS to BEC.

\subsection{ Pairing length }

\begin{figure}[tbp]
\includegraphics[width=8cm]{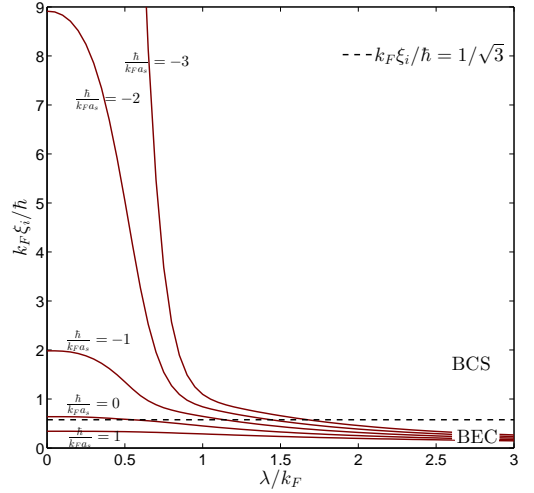} %
\caption{
From the BCS side at $ \lambda=0 $,  the pairing length $\protect\xi _{x}=\protect\xi _{y}=\protect\xi _{z}$ of 3d Weyl SOC decreases quickly and monotonically as the
$\protect\lambda $ increases and drop below the dashed line. It describe precisely the new BCS to BEC crossover driven by the SOC strength.
The dashed line is a contrasting line
where $k_{F}\protect\xi _{0}=\hbar $ (for each component, $k_{F}\protect\xi %
_{0i}=\hbar /\protect\sqrt{3}$). From the BEC side at $ \lambda=0 $, the effects of SOC are small.}
\label{fig7}
\end{figure}

\begin{figure}[tbp]
\includegraphics[width=8cm]{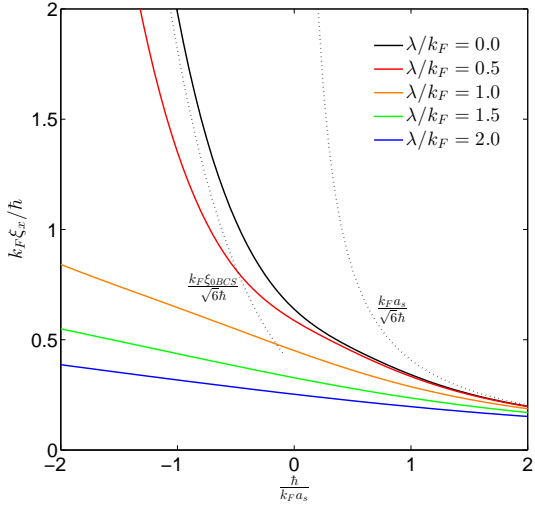}
\caption{ The pairing lengths $ \xi_i $ of 3d Weyl SOC at a fixed SOC versus the scattering length. Different colors stand for different SOC strengths.
The dark dotted line on the left (right) is its BCS (BEC) limit $  \frac{1}{\sqrt{6}} k_F \xi_{0BCS}=\frac{1}{ 8 \sqrt{6} e^{2} } e^{-\frac{\pi \hbar }{2k_{F}a_{s}}} $ ($ \frac{ k_F a_s }{ \sqrt{6} } $) at $ \lambda=0 $. On the BCS side, the SOC effects are dramatic, but on the BEC side,
the SOC effects are small, all curves converge to the right dotted line  $ \frac{ k_F a_s }{ \sqrt{6} } $ from below.
          Compare with Fig.3.  }
\label{fig8}
\end{figure}

   The wavefunction stays the same as Eqn.\ref{3drashbabcs} with the corresponding  $\xi_{\boldsymbol{p}\pm }$ and $E_{\boldsymbol{p}\pm }$
   defined above.  Similar to Sec.III-B, we can determine the singlet pairing amplitude:
\begin{eqnarray*}
\psi _{\substack{ \uparrow \downarrow  \\ \downarrow \uparrow }}\left(
\boldsymbol{p}\right) &=&\frac{\left\langle BCS\right\vert c_{\boldsymbol{p}%
\uparrow }^{\dagger }c_{-\boldsymbol{p}\downarrow }^{\dagger }\left\vert
BCS\right\rangle }{\left\langle BCS|BCS\right\rangle } \\
&=&\mp \frac{1}{2}\left(\frac{w_{\boldsymbol{p}+}}{1+w_{\boldsymbol{p}+}^{2}%
}+\frac{w_{\boldsymbol{p}-}}{1+w_{\boldsymbol{p}-}^{2}}\right)  \nonumber   \\
& - & \frac{1}{2}%
\left(\frac{w_{\boldsymbol{p}+}}{1+w_{\boldsymbol{p}+}^{2}}-\frac{w_{%
\boldsymbol{p}-}}{1+w_{\boldsymbol{p}-}^{2}}\right) \frac{p_{z}}{p},
\end{eqnarray*}%
    and triplet pairing amplitude:
\begin{eqnarray*}
\psi _{\uparrow \uparrow }\left(\boldsymbol{p}\right) &=&\frac{\left\langle
BCS\right\vert c_{\boldsymbol{p}\uparrow }^{\dagger }c_{-\boldsymbol{p}%
\uparrow }^{\dagger }\left\vert BCS\right\rangle }{\left\langle
BCS|BCS\right\rangle } \\
&=&\frac{1}{2}\left(\frac{w_{\boldsymbol{p}+}}{1+w_{\boldsymbol{p}+}^{2}}-%
\frac{w_{\boldsymbol{p}-}}{1+w_{\boldsymbol{p}-}^{2}}\right) \frac{p_{\perp }%
}{p}e^{i\varphi _{\boldsymbol{p}}}=-\psi _{\downarrow \downarrow }^{\ast
}\left(\boldsymbol{p}\right) ,
\end{eqnarray*}%
where $w_{\boldsymbol{p}\pm }=\sqrt{\frac{E_{\boldsymbol{p}\pm }-\xi _{%
\boldsymbol{p}\pm }}{E_{\boldsymbol{p}\pm }+\xi _{\boldsymbol{p}\pm }}}$.

The pairing length $ \xi_{i} $ can be calculated using Eqn.\ref{xii} and
is shown in Fig.7. As the Weyl SOC strength increases, in the BCS side,  the pairing length along any
direction decreases monotonically and quickly, then drop below the dashed line.
In the BEC side, the effects of the SOC strength are quite small.
This is the most direct evidence that the Weyl  SOC can also drive a crossover from
BCS to BEC and can be directly detected by the MIT type of experiment \cite{mitdiss}.

The pairing lengths  versus the scattering length is shown in Fig.8 which is complementary to Fig.7

\subsection{ Cooper pair size }

\begin{figure}[tbp]
\includegraphics[width=8cm]{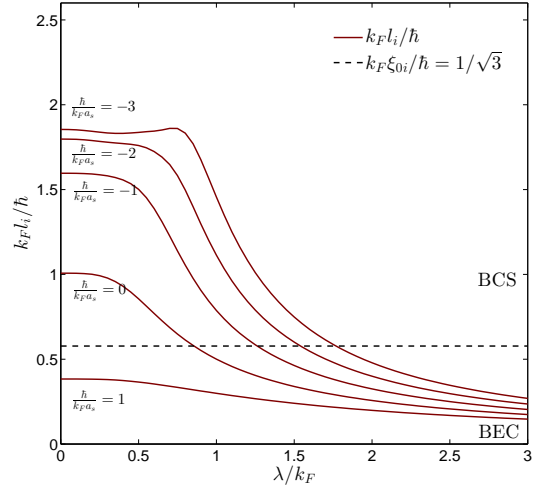}
\caption{  The Cooper pair size $l_{x}=l_{y}=l_{z}$ of 3d Weyl SOC as a
function of $\protect\lambda $. Note its non-monotonic behavior at the BCS side. The SOC effects on the BEC side are small. }
\label{fig9}
\end{figure}

\begin{figure}[tbp]
\includegraphics[width=8cm]{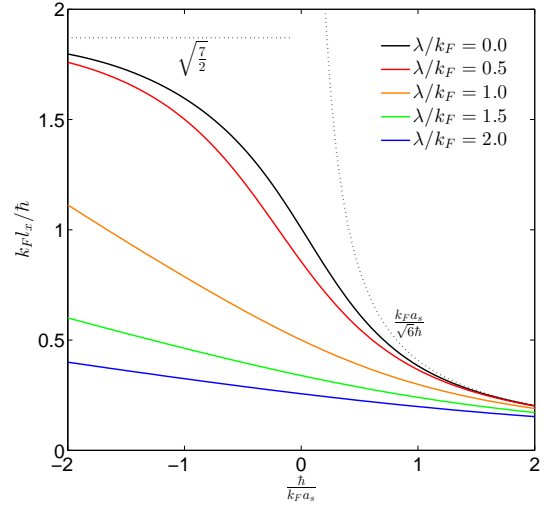}
\caption{ The Cooper pair size $ l_i $ of 3d Weyl SOC at a fixed SOC versus the scattering length. Different colors stand for different SOC strengths.
The dark dotted line on the left (right) is its BCS (BEC) limit $ \sqrt{7/2} $ ($ \frac{ k_F a_s }{ \sqrt{6} } $) at $ \lambda=0 $.
On the BCS side, the SOC effects are dramatic, but on the BEC side,
the SOC effects are small, all curves converge to the right dotted line  $ \frac{ k_F a_s }{ \sqrt{6} } $ from below.
Compare with Fig.5.}
\label{fig10}
\end{figure}

 The Cooper pair wave function takes the same form as
Eqn.\ref{3drashbacp} in the second quantized form and
Eqn.\ref{3drashbacpfirst}  in the first quantized form with the corresponding  $\xi_{\boldsymbol{p}\pm }$ and $E_{\boldsymbol{p}\pm }$
defined above. All the components can be written as:%
\begin{eqnarray}
g_{\uparrow \downarrow }\left(\boldsymbol{p}\right) &=&-\frac{1}{2}\left(
w_{\boldsymbol{p}+}+w_{\boldsymbol{p}-}\right) -\frac{1}{2}\left(w_{%
\boldsymbol{p}+}-w_{\boldsymbol{p}-}\right) \frac{p_{z}}{p}=-g_{\downarrow
\uparrow }\left(-\boldsymbol{p}\right),  \nonumber \\
g_{\downarrow \uparrow }\left(\boldsymbol{p}\right) &=&\frac{1}{2}\left(w_{%
\boldsymbol{p}+}+w_{\boldsymbol{p}-}\right) -\frac{1}{2}\left(w_{%
\boldsymbol{p}+}-w_{\boldsymbol{p}-}\right) \frac{p_{z}}{p},  \nonumber \\
g_{\uparrow \uparrow }\left(\boldsymbol{p}\right) &=&\frac{1}{2}\left(w_{%
\boldsymbol{p}+}-w_{\boldsymbol{p}-}\right) \frac{p_{\perp }}{p}e^{-i\varphi
_{\boldsymbol{p}}}=-g_{\downarrow \downarrow }^{\ast }\left(\boldsymbol{p}%
\right).
\label{3disotropicwf2}
\end{eqnarray}%

  The Cooper-pair size can be evaluated using Eqn.\ref{3drashbal} and  plotted in Fig.9.
  Its non-monotonic behaviors at the BCS side indicate it may not be a good quantity to characterize the crossover.

  The Cooper pair sizes versus the scattering length is shown in Fig.10 which is complementary to Fig.9.

\subsection{ Contrast with the two-body wavefunctions }

 To explore the relations between the many-body wavefunctions or the ``Cooper-pair" wavefunction studied in this section and the two body
 wavefunctions in \cite{twobody}, it is convenient to introduce the spin eigenstate along the momentum
 $ \frac{\boldsymbol{p}}{p}=(\sin \theta \cos \varphi, \sin \theta \sin \varphi, \cos \theta) $:
\begin{eqnarray*}
\left\vert \uparrow \right\rangle _{\boldsymbol{p}} &=&e^{-i\frac{\varphi }{2%
}}\cos \frac{\theta }{2}\left\vert \uparrow \right\rangle +e^{i\frac{\varphi
}{2}}\sin \frac{\theta }{2}\left\vert \downarrow \right\rangle, \\
\left\vert \downarrow \right\rangle _{\boldsymbol{p}} &=&e^{-i\frac{\varphi
}{2}}\sin \frac{\theta }{2}\left\vert \uparrow \right\rangle -e^{i\frac{%
\varphi }{2}}\cos \frac{\theta }{2}\left\vert \downarrow \right\rangle,
\end{eqnarray*}%
then to express the many-body wavefunctions in terms of the spin eigenstates along the momentum
 $ \vec{p} $:
\begin{eqnarray}
&&g_{\uparrow \downarrow }\left(\boldsymbol{p}\right) \left\vert \uparrow
\downarrow \right\rangle +g_{\downarrow \uparrow }\left(\boldsymbol{p}%
\right) \left\vert \downarrow \uparrow \right\rangle +g_{\uparrow \uparrow
}\left(\boldsymbol{p}\right) \left\vert \uparrow \uparrow \right\rangle
+g_{\downarrow \downarrow }\left(\boldsymbol{p}\right) \left\vert
\downarrow \downarrow \right\rangle     \nonumber \\
&=&-\frac{1}{2}\left(w_{p+}+w_{p-}\right) \left(\left \vert \uparrow
\downarrow \right \rangle -\left \vert \downarrow \uparrow \right \rangle
\right) + \frac{1}{2}\left(w_{p+}-w_{p-}\right) \times  \nonumber  \\
& & \left[ \frac{p_{x}-ip_{y}}{p}\left \vert \uparrow \uparrow
\right \rangle -\frac{p_{z}}{p}(\left \vert
\uparrow \downarrow \right \rangle + \left \vert \downarrow
\uparrow \right \rangle) -\frac{p_{x}+ip_{y}}{p}\left \vert \downarrow \downarrow
\right \rangle \right]     \nonumber   \\
&=&-\frac{1}{2}\left(w_{p+}+w_{p-}\right) \left(\left \vert \uparrow
\downarrow \right \rangle -\left \vert \downarrow \uparrow \right \rangle
\right)    \nonumber   \\
 & + & \frac{1}{2}\left(w_{p+}-w_{p-}\right)    \left(\left \vert \uparrow
\downarrow \right \rangle _{\boldsymbol{p}}+\left \vert \downarrow \uparrow
\right \rangle _{\boldsymbol{p}}\right)   \nonumber  \\
&\equiv &g_{a}\left(p\right) \left(\left\vert \uparrow \downarrow
\right\rangle -\left\vert \downarrow \uparrow \right\rangle \right)
+g_{s}\left(p\right) \left(\left\vert \uparrow \downarrow \right\rangle _{%
\boldsymbol{p}}+\left\vert \downarrow \uparrow \right\rangle _{\boldsymbol{p}%
}\right),  \label{3disotropicwf3}
\end{eqnarray}%
 where the components $g_{a}\left(p\right) $ and $g_{s}\left(
p\right) $ are independent on direction of $\boldsymbol{p}$ (i.e. $\theta $
and $\varphi $).  Compared to Eqn.\ref{singlet0}, one can see that there are three extra $ p_x \pm i p_y $ and $ p_z $ pairing components\cite{legg}
similar to the $ B $-phase of superfluid $^{3} He $. This fact should be contrasted to Eqn.\ref{3drashbacpfirst}  where  there are only
two extra equal-spin $ p_x \pm i p_y $  pairing components\cite{legg} similar to the $ A $-phase of superfluid $^{3} He $.

Fourier transforming the ``Cooper-pair" wavefunction Eqn.\ref%
{3disotropicwf3} to real space and comparing with the two body wavefunction in the spherical case in \cite{twobody},
we find that they have the same symmetry.
In fact, a similar relation between the wavefunctions (or order parameters) in real space and those in the helicity momentum basis
were derived for magnetic transitions in repulsively interacting Fermi gas \cite{replusive2}.

\section{2D Fermi gas with a Rashba SOC}

\begin{figure}
\includegraphics[width=7cm]{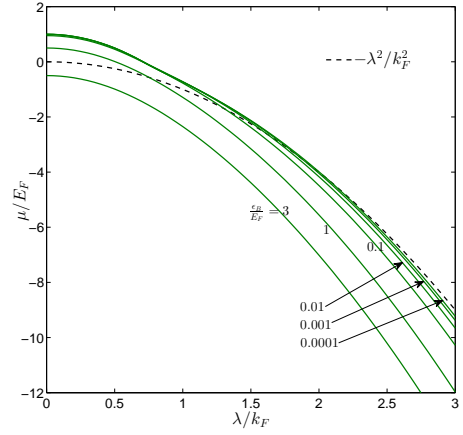}
\caption{ The chemical potential $\protect\mu $ versus $
\protect\lambda $ in 2D isotropic SOC system for different scattering lengths.
The dashed line is the minimum energy of the single particle $\protect\mu _{0}=-\frac{\protect\lambda ^{2}}{2m} $.
On the BCS side, as $\protect\lambda $
increases, the $ \mu $ drops below  $ \protect\mu _{0}$ indicating a crossover from a BCS to
BEC crossover.  }
\label{fig11}
\end{figure}

   A 2D Rashba SOC term can be written as
\begin{equation}
V_{2d-ra}=\frac{\lambda }{m}\underset{\boldsymbol{p}}{\sum }p\left[
e^{-i\varphi _{\boldsymbol{p}}}c_{\boldsymbol{p}\uparrow }^{\dagger }c_{%
\boldsymbol{p}\downarrow }+e^{i\varphi _{\boldsymbol{p}}}c_{\boldsymbol{p}%
\downarrow }^{\dagger }c_{\boldsymbol{p}\uparrow }\right],
\label{2disotropic}
\end{equation}%
where $\lambda $\ is the strength of SOC, $p=\sqrt{p_{x}^{2}+p_{y}^{2}}$ and
$\varphi _{\boldsymbol{p}}=Arg\left(p_{x}+ip_{y}\right) $.
Note that the space is 2d, but the spin is still $ SU(2) $ with the 3 generators.

The BCS theory in two dimension has been studied by several works\cite{xi,xi2} with different focus.
The calculations are similar to the 3d Rashba case in Sec.III with
the momentum $ \vec{p} $ confined to be the 2d momentum $ \vec{p}_{\perp} $,
or similar to the 3d Weyl case in Sec.IV by setting $ p_z =0 $.
Eqn.\ref{h03dra},\ref{helicitybase}, \ref{3drashbah}, \ref{3drashbab} follow.
The two self-consistent equations Eqn.\ref{3dconsistent} also hold with the crucial difference
that the interaction need to be regularized by a bound state energy $ \epsilon_B $ at 2d,
instead of a scattering length $ a_s $ in 3d:
$ \frac{1}{g}=-\frac{1}{V}\underset{\boldsymbol{p}}{
\sum }\frac{1}{2\epsilon _{\boldsymbol{p}}+\epsilon _{B}}$.
Solving them leads to the chemical potential $ \mu $ shown in Fig.11.

\subsection{ Pairing length }

\begin{figure}
\includegraphics[width=7cm]{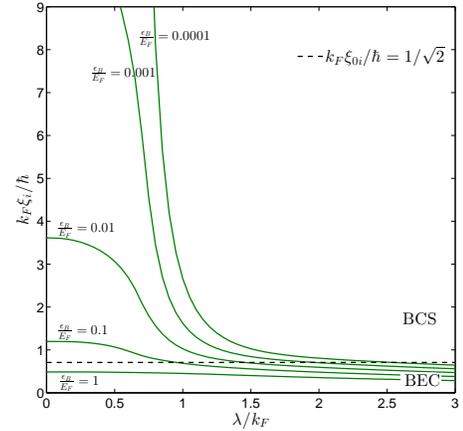}
\caption{  The 2d pairing length defined in Eqn.\ref{xii} ($\protect\xi _{x}=\protect\xi _{y}$) as a
function of $\protect\lambda $. At the BCS side, it decreases quickly and monotonically as the
$\protect\lambda $ increases and drop below the dashed line. It describe precisely the new BCS to BEC crossover driven by the SOC strength.
The dashed line is a guidance line
where $k_{F}\protect\xi_0 =\hbar $ (averagely, $k_F\protect\xi_{0i}=\hbar/
\protect\sqrt{2}$). At the BEC side, the effects of SOC are small. }
\label{fig12}
\end{figure}

\begin{figure}[tbp]
\includegraphics[width=7cm]{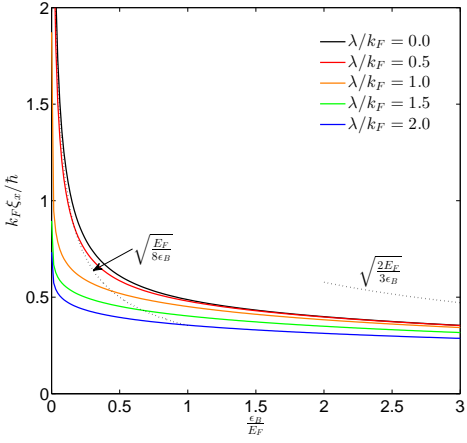}
\caption{ The 2d pairing lengths $\xi_x=\xi_y $ at a fixed SOC versus the scattering length. Different colors stand for different SOC strengths.
 The black dotted line on the left (right) is its BCS (BEC) limit $ \sqrt{\frac{ E_F}{8 \epsilon_B}} $
 ($ \sqrt{ \frac{ 2 E_F}{3 \epsilon_B}} $) at $ \lambda=0 $. On the BCS side, the SOC effects are dramatic, but on the BEC side,
the SOC effects are small, all curves converge to the right dotted line  $  \sqrt{ \frac{ 2 E_F}{3 \epsilon_B}} $ from below. }
\label{fig13}
\end{figure}

When calculating the pairing length, Eqns.\ref{3drashbabcs}, \ref{3dsinglet}, \ref{3dtriplet}
still hold. For $\lambda \neq 0$, only numerical results are available and are shown in Fig.12.
We can see that in the BCS limit at $ \lambda=0 $, as the strength of SOC increases for a fixed
interaction strength $\frac{\epsilon _{B}}{E_{F}}$, the pair size decreases monotonically and sharply, then below the
the reference line. Here, we also plot a
reference line by taking $k_{F}\xi _{0}=\hbar $ (for each component, $k_{F}\xi _{0i}=
\frac{\hbar }{\sqrt{2}}$) to qualitatively observe the BCS to BEC crossover behavior.
In the BEC limit, the effects of SOC are small.

Setting $ \lambda=0 $, one can easily solve the
self-consistent equations and find $\mu =E_{F}-\frac{\left\vert \epsilon
_{B}\right\vert }{2}$ and $\Delta =\sqrt{2\left\vert \epsilon
_{B}\right\vert E_{F}}$. When $\lambda =0$, Eqn.\ref{3drashbasize0} at $ d=3 $ should be replaced by \cite{xi,xi2}:
\begin{eqnarray}
\xi _{x,y}^{2} (\lambda=0) &=&\frac{\hbar ^{2}4\int dpp^{d+1}\frac{\xi _{\boldsymbol{p}%
}^{2}}{E_{\boldsymbol{p}}^{6}}}{\left(2m\right) ^{2}d\int dp\frac{p^{d-1}}{%
E_{\boldsymbol{p}}^{2}}}  \nonumber \\
&=&\frac{\hbar ^{2}}{4\left(2m\Delta \right) }\left(\eta +\frac{\eta ^{2}+2%
}{\eta ^{2}+1}\frac{1}{\frac{\pi }{2}+\arctan \eta }\right),
\label{2disotropicsize0}
\end{eqnarray}%
where $\eta =\frac{\mu }{\Delta }$. Because of different dimensions, this
analytical expression is very different from Eqn.\ref{3drashbasize0} in 3d.
In the BCS limit (i.e. $\frac{\epsilon _{B}}{E_{F}}
\longrightarrow 0$), $\eta =\frac{1}{\sqrt{2}}\sqrt{\frac{E_{F}}{\left\vert
\epsilon _{B}\right\vert }}\longrightarrow \infty $,
$   k_F \xi _{x,y} \longrightarrow \sqrt{ \frac{ E_F }{8 \left\vert \epsilon _{B}\right\vert } } $ which diverges.
In the BEC limit (i.e. $\frac{\left\vert \epsilon _{B}\right\vert }{E_{F}}
\longrightarrow \infty $), $\mu =-\frac{\left\vert \epsilon _{B}\right\vert
}{2}$ and $\eta =-\frac{1}{2\sqrt{2}}\sqrt{\frac{\left\vert \epsilon
_{B}\right\vert }{E_{F}}}\longrightarrow -\infty $,
$ k_F \xi _{x,y} \rightarrow \sqrt{ \frac{3}{2} \frac{ E_F }{\left\vert \epsilon _{B}\right\vert }} $.

Shown in Fig.13 is the pairing length versus the scattering length which is complementary to Fig.12.

\subsection{ Cooper pair size }

\begin{figure}
\includegraphics[width=7cm]{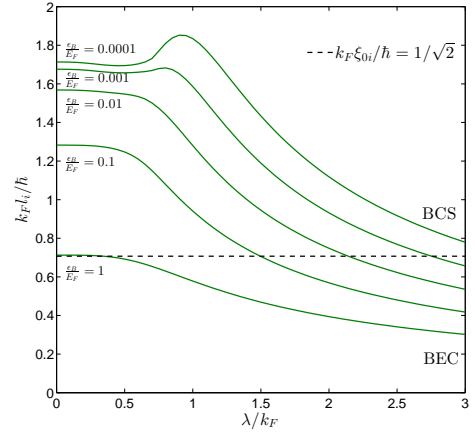}
\caption{ The 2d Cooper pair size  $l _{x}=l _{y}$  as a function of $\protect\lambda $.  Note its non-monotonic behavior at the BCS side. }
\label{fig14}
\end{figure}

\begin{figure}[tbp]
\includegraphics[width=7cm]{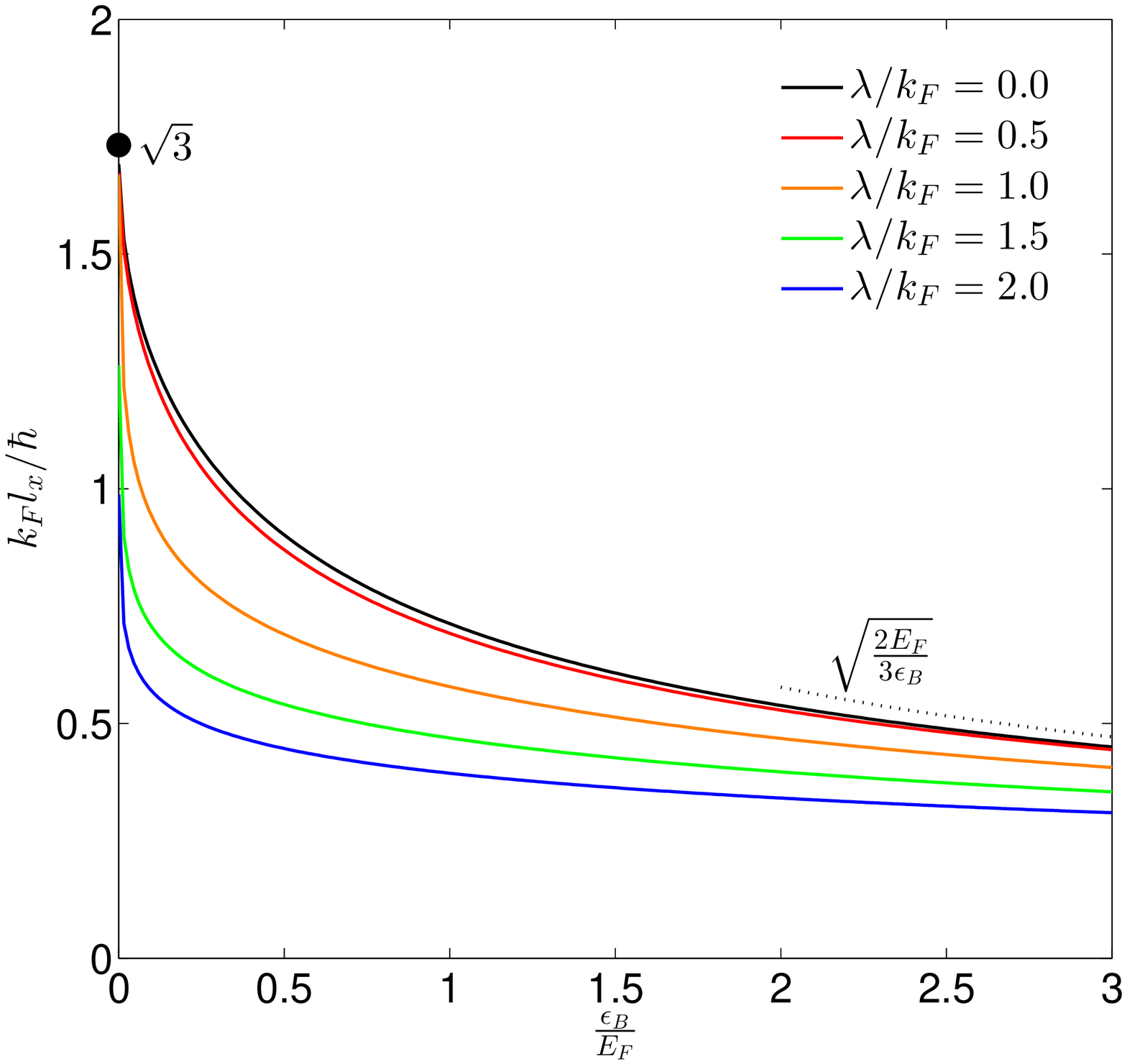}
\caption{ The 2d Cooper pair size $ l_x=l_y $ at a fixed SOC versus the scattering length. Different colors stand for different SOC strengths.
The black dotted line on the left (right) is its BCS (BEC) limit $ \sqrt{3} $ ($ \sqrt{ \frac{2 E_F}{3\epsilon_B} } $) at $ \lambda=0 $.
On the BCS side, the SOC effects are dramatic, but on the BEC side,
the SOC effects are small, all curves converge to the right dotted line  $  \sqrt{ \frac{ 2 E_F}{3 \epsilon_B}} $ from below.  }
\label{fig15}
\end{figure}

When calculating the Cooper-pair size, Eqns. \ref{3drashbacp}, \ref{3drashbawf2}, \ref{3drashbacpfirst}, \ref{3drashbal}
still hold. Shown in Fig.14 is how the Cooper-pair size changes with $ \lambda $.
Once more, its  non-monotonic behaviors  at the BCS side indicate it may not be a good quantity to characterize the BCS to BEC  crossover.

 When $\lambda =0$, Eqn.\ref{3dnosocl} at $ d=3 $ should be replaced by:
\begin{widetext}
\begin{equation}
l_{i}^{2}(\lambda=0) =\frac{3\hbar ^{2}\left[ \ln 2-\frac{1}{2}-\eta \left(\frac{\pi }{2%
}+\arctan \eta \right) -\ln \left(1-\frac{\eta }{\sqrt{\eta ^{2}+1}}\right)
+\eta \left(\eta +\sqrt{\eta ^{2}+1}\right) \right] }{2m\Delta \left[ \eta
^{3}+\left(\eta ^{2}+1\right) ^{\frac{3}{2}}+\frac{3}{2}\eta \right] },
\label{2dnosocl}
\end{equation}
\end{widetext}
where $\eta =\frac{\mu }{\Delta }$. In the BCS limit (i.e.$
\frac{\epsilon _{B}}{E_{F}}\longrightarrow 0$), one get $
l_{x,y} \longrightarrow \sqrt{3} \frac{\hbar }{k_{F} }$ which is nothing but the inter-particle distance, so it goes
to a finite value, in sharp contrast to the pairing length which diverges.
In fact, $ l/\xi \sim \Delta_0/\epsilon_F  \rightarrow 0 $ in the BCS limit.
In the BEC limit (i.e.$ \frac{\left\vert \epsilon _{B}\right\vert }{E_{F}}
\longrightarrow \infty $), one find
$  k_F l_{x,y} =\sqrt{\frac{ 2 E_F }{ 3 \left\vert \epsilon _{B}\right\vert } } $ which is identical to
$  k_F \xi _{x,y} $ in the BEC limit as it should be.

Shown in Fig.15 is the Cooper pair size at various fixed SOC strengths  versus the bound state energies $ \epsilon_B/E_F $
which is complementary to Fig.14.

\section{ Applications to 2d superconductor and semi-conductor systems }

   In various 3d condensed matter systems \cite{ahe,kitpconf}, the 3d SOC usually take
   $ \lambda (\vec{k} \times \vec{\sigma}) \cdot \nabla V $  which is quite different form than Weyl or Rashba
   form studied in Sec.III and IV by keeping the inversion symmetry. It may be interesting to see if such a 3d inversion symmetric SOC
   can also drive a BCS to BEC crossover.

   Eqn.\ref{hfermion} with the 2D Rashba SOC term Eqn.\ref{2disotropic} may also describe 2d bright exciton  with total angular momentum
   $ J= \pm 1 $ in electron-hole semiconductor bilayer systems  and electron pairings in 2d  non-centrosymmetric superconductors\cite{socsemi,niu,wu}.
   It was known that in a 2d semiconductor electron gas, the 2d Rashiba SOC strength depends on the electric field, presence of adatoms at
   the boundary, atomic weight and atomic shells involved \cite{socsemi,Kane,Zhang}.
   In the surface of non-centrosymmetric superconductors, the strong near surface electric fields lead to a 2d Rashba SOC
   quite similar to the 2d superconducting fullerene and polyacene crystals in the field-effect-transistor geometry \cite{socsemi}.
   So the 2d Rashba SOC strength in the two condensed matter systems can also be tuned by adjusting various surface geometries.
   So the results achieved on the BCS to BEC crossover tuned by the 2d Rashba strength in the section V should also apply to these
   condensed matter systems.
    In Ref.\cite{power,squ,excitoncorr,exciton2}, the authors ignored the spins of the electrons and holes, therefore also the
    possible Rashba SOC. As shown at the end of \cite{excitoncorr}, the bright excitons couple to the one photon process with the polarization $ \sigma=\pm $.
    By incorporating  the coupling between the 2d bright excitons subject to the 2d Rashba SOC studied in Sec.V  and the 3d emitted photons with the two polarizations,  it is interesting study how the emitted photon characteristics change
    across the new BCS to BEC crossover driven by the 2d Rashiba SOC.

\section{ Discussions and Conclusions }

    The new BCS to BEC crossover  driven by the SOC strength has been studied by previous authors from the overlap
    between  ``Cooper-pair wavefunction" and two body wavefunction \cite{crossover},
    also from the ``Cooper-pair size" right at the Feshbach resonance \cite{zhai}.
    In this paper, we investigate the new BCS to BEC crossover from fundamental and physical points of view.
    At the mean field level, we studied the dependence of chemical potential, pairing length, Cooper-pair size
    on the SOC strength for three kinds of Fermi gases with 3d Rashba, 3d Weyl and 2d Rashba SOC respectively.
    We explicitly demonstrated the new BCS to BEC crossover
    driven by the SOC strength in all the three cases by  monitoring the monotonic decreasing of chemical potential and the pairing length.
    We show that the most relevant wavefunction is the many body wave function instead of the ``Cooper-pair wavefunction" or two body wavefunction,
    the most relevant length is the pairing length instead of the``Cooper-pair size" or the two-body bound state size.
    Among the three lengths, only the pairing length is the experimentally detectable length.

    We can summarize the main differences among the pairing length, the Cooper-pair size and the two-body size in the following:
    In the absence of SOC, in the BCS limit, the pairing length goes to the coherence length $ \xi(\lambda=0) = \hbar v_F/\Delta_0 $, while
    the Cooper-pair size goes to the inter-particle distance $ l(\lambda=0) \sim  1/k_F $. Their ratio
    $ l(\lambda=0)/\xi(\lambda=0) \sim \Delta_0/\epsilon_F $.
    For conventional superconductors \cite{superbook}, $ l(\lambda=0)/\xi(\lambda=0) \sim 10^{-4} $,
    so they are well inside the BCS limit. The BCS mean field
    theory work well, quantum and classical fluctuation effects can be neglected except very close to the critical transitions at finite temperatures.
    For high temperature superconductors \cite{legg,hightc}, $ l(\lambda=0)/\xi(\lambda=0) \sim 10^{-1} $, so they are quite close to BCS to BEC crossover,
    but still in the BCS limit with well defined Fermi surface. So quantum and classical fluctuation effects can not be ignored \cite{legg}.
    In the BEC limit, they both get to the two-body bound state size, therefore $ l(\lambda=0)/\xi(\lambda=0) \sim 1 $.
    The results on $ \xi $ in Eqn.\ref{3drashbasize0} at 3d \cite{melo} and Eqn.\ref{2disotropicsize0} at 2d \cite{xi,xi2}  are not new,
    but the results on $ l $ in Eqn.\ref {3dnosocl} at 3d and Eqn.\ref{2dnosocl} are new and show completely different behaviors than  $ \xi $.
    It is very instructive to compare the two different length scales.
    In the presence of SOC, on the BCS side, the pairing length $ \xi_i(\lambda), i=x,y,z $  decrease monotonically and  quickly move
    into the BEC regime, so can be used to characterize the BCS to BEC crossover quantitatively.
    Furthermore it can be detected by RF dissociation spectra experiment.
    While $ l_i(\lambda) $ shows non-monotonic behaviors, so can not be used to characterize the BCS to BEC crossover even qualitatively.
    Furthermore it is not an experimentally measurable quantity.

    In a future publication, we will compute the fluctuation corrections to the mean filed theory results on the pairing length achieved in this paper.
    One can achieve the goal by calculating the fermion pairing correlation function Eqn.\ref{ggspin} using $1/N $ expansion \cite{dicke1,dicke2}.
    It was known the quantum fluctuation effects are important near the BCS to BEC regime.
    It may also be interesting to extend the zero temperature results on the pairing length
    to finite temperatures whose effects are especially important to 2d Rashba systems studied in Sec.V and VI.
    However, it is not known how to extend the concepts of Cooper-pair size defined in Eqn.\ref{li} beyond mean field results and to finite temperatures.
    Above all, its definition is based on the explicit form of the mean field state. Therefore, the pairing length is a much more robust concept
    than the Cooper-pair size. It is also a experimentally measurable quantity through radio-frequency dissociation spectra.
    Of course, the two-body wavefunction is defined only for two fermions, can not be used to study a many body system anyway.
    The Cooper-pair size has been evaluated  at the mean field through the topological transition in \cite{read}. As demonstrated in this paper,
    the pairing length show quite different behaviors than the Cooper-pair size,
    it maybe useful to study the pairing length through various topological transitions driven by the Zeeman field \cite{topo2dra,topo3d,luo}.

{\bf Acknowledgements }

We thank Fadi Sun for helpful discussions.
YY and JY are supported by NSF-DMR-1161497, NSFC-11174210, Beijing Municipal Commission of Education under Grant No. PHR201107121.
WL was supported by the NKBRSFC under grants Nos. 2011CB921502, 2012CB821305, NSFC under grants Nos. 61227902, 61378017, 11311120053.


\begin{thebibliography}{9}
\bibitem{explain}  As shown in Sec.II-B, the reason to put a quotation mark on both ``Cooper-pair wavefunction" and ``Cooper-pair size"
is that they are not really physical, so may not be really understood as the Cooper-pair wavefunction and  Cooper-pair size.
For the illustration purposes,  they are evaluated here to compare with the physical pairing length and two-body bound state.
Furthermore, several authors \cite{read,zhai}  also evaluated this un-physical length in different contexts.
For notation simplicties, we dropped the quotation mark in the following.
\bibitem{factor} Note that $ \xi= \sqrt{ \xi^2_x + \xi^2_y + \xi^2_z } = \frac{ a_s }{ \sqrt{2} } $ which is exactly
   the size of the two-body bound state in Eqn.\ref{bound}.
\bibitem{uncover} In Ref.\cite{zhai}, only the $ l_i $ at the resonance $ \hbar/k_F a_s = 0 $ was calculated.
 Our curve in Fig.4 at the resonance recovers their results. But as indicated in the text,
 the non-monotonic behaviors of $ l_i $ discovered in the Fig.4 can only be seen in the BCS side $ a_s < 0 $. Of course, as indicated in the text,
 the $ l_i $ can not be used to characetrize the BCS to BEC crossover and is not an experimental measurable quantity anyway.
\end{thebibliography}
\end{document}